\documentclass[11pt,a4paper]{article}

\usepackage{jcappub}
\makeatletter
\def\@fpheader{\relax}

\usepackage{appendix}

\usepackage{graphicx}
\usepackage{amsmath}
\usepackage{amssymb}

\newcommand{\be}{\begin{equation}}
\newcommand{\ee}{\end{equation}}
\newcommand{\beq}{\begin{equation}}
\newcommand{\eeq}{\end{equation}}

\newcommand{\vect}[1]{\boldsymbol{\rm #1}}

\newcommand{\Msun}{{\rm M}_\odot}

\title{Implications of hydrodynamical simulations for the interpretation of direct dark matter searches}

\author{Nassim Bozorgnia}
\author{and Gianfranco Bertone}
\affiliation{GRAPPA, Institute for Theoretical Physics Amsterdam, \\
and Delta Institute for Theoretical Physics, University of Amsterdam, \\
Science Park 904, 1098 XH Amsterdam, The Netherlands} 

\emailAdd{n.bozorgnia@uva.nl}
\emailAdd{g.bertone@uva.nl}

\abstract{
In recent years, realistic hydrodynamical simulations of galaxies like the Milky Way have become available, enabling a reliable estimate of the dark matter density and velocity distribution in the Solar neighborhood. We review here the status of hydrodynamical simulations and their implications for the interpretation of direct dark matter searches. We focus in particular on: the criteria to identify Milky Way-like galaxies; the impact of baryonic physics on the dark matter velocity distribution; the possible presence of substructures like clumps, streams, or dark disks; and on the implications for the direct detection of dark matter with standard and non-standard interactions.}


\begin{document}

\maketitle

\section{Introduction}	

A large body of observational evidence from the scale of dwarf galaxies to cosmological scales indicates that 27\% of the mass-energy content of the Universe is in the form of dark matter (DM)~\cite{Ade:2015xua,Bertone:2010zza}. Galaxies like our own are believed to be embedded in a DM halo which extends to more than ten times the scale of the visible galaxy. Broad efforts to identify the particle nature of DM during the last decade have brought together the particle physics, cosmology and astrophysics communities, and may ultimately enable us to constrain the nature of DM by combining the results of cosmological simulations, astrophysical observations, and particle DM searches. 

In standard cosmology, the DM problem requires new physics beyond the Standard Model of particles physics, since none of the known particles have the required properties to be the DM particle. One of the most extensively studied class of DM candidates is that of Weakly Interacting Massive Particles (WIMPs) which can be searched for via direct, indirect and accelerator searches~\cite{Bertone:2010zza,Jungman:1995df,Bergstrom00,Bertone05}. 

Direct DM detection experiments search for WIMPs by measuring the small recoil energy of a target nucleus deposited in an underground detector due to a collision with a WIMP. Several direct detection experiments are running around the globe, and have set stringent constraints on the DM-nucleon interaction cross section. For more than a decade the DAMA experiment has reported a hint for a DM signal~\cite{Bernabei:2016ivu} which is in strong tension with the results of other direct detection experiments. The LUX (Large Underground Xenon) experiment currently sets the strongest exclusion limits in the DM mass and spin-independent DM-nucleon cross section plane for large ($>6$ GeV) DM masses~\cite{Akerib:2016vxi}. 

The interpretation of direct detection results is complicated due to large uncertainties in the astrophysical distribution of DM in the Solar neighborhood. When presenting results in the DM mass and scattering cross section plane, it is usually assumed that the DM distribution in the Galaxy is described by the so-called Standard Halo Model (SHM)~\cite{Drukier:1986tm}: an isothermal sphere of DM, with an isotropic Maxwell-Boltzmann velocity distribution in the Galactic rest frame with the most probable (peak) speed equal to the local circular speed. The DM particles are assumed to be in hydrostatic equilibrium, where the random velocity of the DM particles in the halo provides a collisionless pressure which balances the gravitational potential of the galaxy, and supports the halo from collapsing~\cite{Drukier:1986tm}. 

The SHM provides however a simplistic description of the DM distribution, and modern estimates rely instead on cosmological simulations of galaxy formation. Using the $\Lambda$CDM cosmological model with parameters inferred from precise cosmic microwave background measurements as the initial conditions, the universe in fact can be simulated starting at very large redshifts before structure formation. 
We point the reader to Refs.~\cite{Kuhlen:2012ft} and \cite{Frenk:2012ph} for detailed reviews of cosmological simulations.

Earlier N-body  simulations were performed assuming that all the matter content of the universe is in the form of collisionless DM. High resolution DM-only (DMO) simulations predict local DM velocity distributions which deviate substantially from a Maxwellian distribution\cite{Vogelsberger:2008qb, Kuhlen:2009vh}. 
However, DMO simulations have significant systematic uncertainties in their predictions due to neglecting the effect of baryons. Baryons play a significant role in the process of galaxy formation, and are necessary to draw realistic predictions from cosmological simulations.

In recent years, hydrodynamical simulations which include baryonic physics have become possible and realistic. This is mainly due to advances in physical modeling of  baryonic processes, the exponential growth of computing power, and improvement in numerical techniques. Many of the current hydrodynamical simulations are able to reproduce key properties of galaxies, and have achieved significant agreement with observations.

DM substructures in both the spatial and velocity distribution of the Milky Way (MW) halo in the form of DM streams or DM clumps can also affect direct detection event rates. The effect of the DM stream associated with the accretion of the Sagittarius dwarf galaxy on dark matter event rates has been studied in isolated simulations of the MW, and shown to be non-negligible~\cite{Purcell:2012sh}. However, high resolution DMO simulations predict that the local DM distribution is very smooth. The DM density in the Solar neighborhood is at most 15\% different from the mean over the best fit ellipsoidal equidensity contour at the 99.9\% CL~\cite{Vogelsberger:2008qb}. DMO simulations have also found that DM streams in the Solar neighborhood are unlikely to be important~\cite{Vogelsberger:2010gd}. The DM substructures of the MW have recently been studied in hydrodynamical simulations, and it has been found that the presence of baryons causing tidal disruption reduces the DM substructures in the inner parts (within $\sim 10$ kpc) of the MW~\cite{Sawala:2016tlo, Garrison-Kimmel:2017zes}. However higher resolution hydrodynamical simulations are needed to study the DM substructures at the Solar position. 

In this article, we will review the local DM distribution extracted from different hydrodynamical simulations and the implications for DM direct detection. In particular we will discuss the predictions for direct detection of the simulations performed by Ling {\it et al.}~\cite{Ling:2009eh}, Eris~\cite{Kuhlen:2013tra}, NIHAO~\cite{Butsky:2015pya}, EAGLE and APOSTLE~\cite{Bozorgnia:2016ogo}, MaGICC~\cite{Kelso:2016qqj}, and Sloane {\it et al.}~\cite{Sloane:2016kyi}. The paper is structured as  follows. In Section~\ref{Notation} we review the formalism for computing direct detection event rates and the astrophysical inputs relevant for direct detection. In Section~\ref{simulations} we present the details of the different simulations studied in this work. In Section~\ref{IdentifyMW} we discuss the different criteria used by different simulation groups to identify simulated MW analogues. In Section~\ref{DMdensity} we review the local DM density extracted from different simulations, as well as the halo shapes of the MW analogues. In Section~\ref{f(v)} we discuss the local DM velocity distributions of different haloes and how they compare to the Maxwellian velocity distribution. The possibility for the existence of a dark disk in different simulations is also studied. In Section~\ref{Implications} we review the implications for direct detection and present an analysis of direct detection data using the DM distribution from different simulations. In Section~\ref{Nonstandard} we comment on non-standard DM-nucleus interactions, and we conclude in Section~\ref{conclusions}.


\section{Dark matter direct detection}
\label{Notation}

\subsection{Event rate in DM direct detection}

Consider the elastic collision of a DM particle $\chi$ with mass $m_\chi$ and a target nucleus with mass $m_A$ and atomic mass number $A$. The energy differential event rate (per unit energy, detector mass, and time) is given by

\beq \label{rate}
\frac{d R}{d E_R} = \frac{\rho_\chi}{m_\chi} \frac{1}{m_A}\int_{v>v_{\rm min}}d^3 v \frac{d\sigma_A}{d{E_R}} v f_{\rm det}(\vect v, t),
\eeq
where $E_R$ is the energy of the recoiling nucleus, $\rho_\chi$ is the local DM density, $d\sigma_A/d E_R$ is the energy differential DM--nucleus scattering cross section, $f_{\rm det}(\vect v, t)$ is the local DM velocity distribution in the detector rest frame normalized to 1, and  $\vect v$ is the relative velocity between DM and the nucleus, while $v\equiv |\vect{v}|$. For a DM particle to deposit a recoil energy $E_R$ in the detector, a minimum speed  $v_{\rm min}$ is required,
\beq v_{\rm min}=\sqrt{\frac{m_A E_R}{2 {\mu_{\chi
        A}^2}}},
\label{eq:vm}
\eeq
where $\mu_{\chi A}=m_\chi m_A/(m_\chi + m_A)$ is the DM--nucleus reduced mass.

For the case of spin-independent DM-nucleus scattering with equal couplings of DM to protons and neutrons, the energy differential DM--nucleus cross section can be written in terms of the spin-independent DM-nucleon scattering cross section, $\sigma_{\rm SI}$, 
\begin{align}
  \frac{d\sigma_A}{dE_R} = \frac{m_A A^2}{2\mu_{\chi p}^2 v^2} {\sigma_{\rm SI}} F^2(E_R) \,.
  \label{eq:dsigmadE}
\end{align}
Here, $\mu_{\chi p}$ is the DM-nucleon reduced mass, and $F(E_R)$ is a form factor taking into account the finite size of the nucleus. 

From Eqs.~\eqref{rate} and \eqref{eq:dsigmadE}, the energy differential event rate can be written as
\beq\label{eq:Reta}
\frac{d R}{d E_R} = \frac{A^2 \sigma_{\rm SI}}{2 m_\chi \mu_{\chi p}^2} \, F^2(E_R) \, \rho_\chi \eta(v_{\rm min}, t),
\eeq
where 
\beq\label{eq:eta} 
\eta(v_{\rm min}, t) \equiv \int_{v > v_{\rm mim}} d^3 v \frac{f_{\rm det}(\vect v, t)}{v} \,,
\eeq
is the halo integral, which together with the local DM density $\rho_\chi$ contain the astrophysical dependence of the event rate.

The time dependence of the differential event rate is due to the time dependence of the velocity of the Earth with respect to the Sun, $\vect v_e(t)$. To find the DM velocity distribution in the detector frame, one has to boost the DM velocity in the Galactic frame by the Sun's velocity with respect to the center of the Galaxy, $\vect v_s$, and the Earth's velocity with respect to the Sun $\vect v_e(t)$, such that, $f_{\rm det}(\vect v, t) = f_{\rm gal}(\vect v + \vect v_s + \vect v_e(t))$.

\subsection{Astrophysical inputs}

The DM density, $\rho_\chi$, and velocity distribution, $f_{\rm det}(\vect v, t)$, at the position of the Sun are the astrophysical inputs entering in the direct detection event rate. To present the results of different direct detection experiments in the plane of the DM-nucleon scattering cross section and the DM mass, and hence to make any predictions for the particle physics nature of DM, an assumption for $\rho_\chi$ and  $f_{\rm det}(\vect v, t)$ is required. 

Notice that the local DM density enters as a normalization in the event rate, while the DM velocity distribution enters the event rate through an integration over DM velocities in the detector (see Eq.~\ref{rate}). Depending on the target nuclei and the energy range probed by a direct detection experiment, the minimum speed, $v_{\rm min}$,  is determined for that particular experiment. Thus, different experiments probe different DM speed ranges and their dependence on the DM velocity distribution varies.

In the SHM, the DM velocity distribution in the Galactic frame is assumed to be a Maxwellian distribution with a peak speed equal to the local circular speed, $v_c$, and truncated at the escape speed, $v_{\rm esc}$, from the Galaxy,
\beq\label{eq:Maxw} 
f_{\rm gal}({\bf v}) = \begin{cases}  N\exp{\left(-{\bf v}^2/v_c^2\right)} & v<v_{\rm esc}\\
0 & v \geq v_{\rm esc} \end{cases}
\eeq
where $N$ is a normalization factor. The local circular speed is usually assumed to be 220 or 230 km$/$s, and the commonly adopted escape speed is 544 km$/$s. 

When adopting the SHM, the uncertainties in the local circular speed and escape speed from the Galaxy are usually neglected. The local circular speed can range from $(220 \pm 20)$~km$/$s to $(279 \pm 33)$~km$/$s~\cite{McMillan:2009yr}. Even if the Maxwellian functional form describes well the local DM velocity distribution, the relationship between its peak speed and the local circular speed can be nontrivial.  The Galactic escape speed at the Solar position found by the RAVE survey is $v_{\rm esc} = 533^{+54}_{-41}$~km$/$s at the 90\% CL. 

The fiducial value for the local DM density in the SHM is 0.3 GeV$/$cm$^3$. There are two main approaches for determining $\rho_\chi$ from observations: local methods, where kinematical data from a nearby population of stars is used to constrain the total Galactic potential, and global methods, which are based on mass modeling the DM and baryonic components of the MW and fits to kinematical data across the whole Galaxy. The recent local and global estimates of the local DM density are in the range $\rho_\chi = (0.2 - 0.8)$ GeV$/$cm$^3$~\cite{Read:2014qva, Catena:2009mf, Weber:2009pt, Salucci:2010qr, McMillan:2011wd, Garbari:2011dh, Iocco:2011jz, Garbari:2011tv, Bovy:2012tw, Zhang:2012rsb, Bovy:2013raa,Pato:2015dua, Silverwood:2015hxa, Huang:2016, McMillan:2016}.


\section{Hydrodynamical simulations}
\label{simulations}

Recently, many realistic hydrodynamical simulations have become possible, and several simulation groups have been able to produce disk galaxies with realistic masses and sizes. In this section we broadly review the hydrodynamical simulations which have been used to extract the DM distribution in MW-like galaxies and study the implications for DM direct detection. We point the reader to Refs.~\cite{Ling:2009eh}, \cite{Kuhlen:2013tra}, \cite{Butsky:2015pya}, \cite{Bozorgnia:2016ogo}, \cite{Kelso:2016qqj}, \cite{Sloane:2016kyi}, and the references therein for the details of each simulation. The parameters of the simulations discussed in this review are presented in Table~\ref{tab:sims}. Notice that each simulation adopts a different method for solving the hydrodynamical equations, a different galaxy formation model, has a different spatial resolution, and a different mass for the DM particle.

The cosmological parameters used as initial conditions in the simulations are given in Table~\ref{tab:CosmoPar}. We expect that the small differences in the cosmological parameters used by different simulation groups would not cause a significant variation among their results. The different hydrodynamical prescriptions and parameters of the simulations (given in Table~\ref{tab:sims}) are expected to cause a much larger variation among the end results.

Some of the most relevant baryonic processes for causing macroscopic changes in a simulated galaxy are gas cooling processes, star formation treatment, and the supernova feedback mechanism. Except for Eris, all the simulations discussed below assume that hydrogen, helium, and metals are present, and include metal line cooling processes.

\begin{table}
\centering
\begin{tabular}{@{}cccccc@{}}
\hline
Simulation & code & $N_{\rm DM}$ & $m_{\rm g}~[\Msun]$ & $m_{\rm DM}~[\Msun]$ & $\epsilon$ [pc] \\ 
\hline
Ling {\it et al.} &{\sc ramses}& $2662$ & -- & $7.46\times10^{5}$ & 200 \\
Eris & {\sc gasoline} & 81213 & $2\times10^{4}$ & $9.80\times10^{4}$ & 124 \\
NIHAO &{\sc EFS-Gasoline2} & -- & $3.16 \times10^{5}$ & $1.74\times10^{6}$ & 931 \\
EAGLE (HR) &{\sc P-Gadget (anarchy)} & 1821--3201 & $2.26\times10^{5}$ & $1.21\times10^{6}$ & 350 \\
APOSTLE (IR) & {\sc P-Gadget (anarchy)}   & 2160, 3024 & $1.3\times10^{5}$ & $5.9\times10^{5}$ & 308 \\          
MaGICC & {\sc gasoline} & 4849, 6541 & $2.2 \times 10^5$ & $1.11\times10^{6}$ & 310 \\
Sloane {\it et al.} & {\sc gasloine}& 5847--7460 & $2.7 \times 10^4$ & $1.5\times10^{5}$ & 174 \\
\hline
\end{tabular}
\caption{Parameters of the simulations discussed in this article. The columns specify the code used to perform the hydrodynamical simulation, the number of DM particles, $N_{\rm DM}$, 
     within the torus region defined for each simulation in Section \ref{SpeedDist} around the Solar circle of the selected haloes, the initial gas particle mass, $m_g$,
     the DM particle mass, $m_{\rm DM}$, and  the Plummer-equivalent
      physical softening length, $\epsilon$. The initial gas particle mass is undefined for the Ling {\it et al.} simulation since a grid code is used in that case. }
\label{tab:sims}
\end{table}

\begin{table}
\centering
\begin{tabular}{@{}ccccccc@{}} 
\hline
Simulation & $\Omega_m$ & $\Omega_\Lambda$ & $\Omega_b$ & $H_0$ [km s$^{-1}$ Mpc$^{-1}$] & $n_s$ & $\sigma_8$ \\
\hline
Ling {\it et al.} & 0.3 & 0.7 & 0.045 & 70 & -- & -- \\
Eris & 0.268 & -- & 0.042 & 73 & 0.96 & 0.76 \\
NIHAO &0.3175 & 0.6825 & 0.0490 & 67.1 & 0.9624 & 0.8344 \\
EAGLE (HR) &0.307 &0.693 &0.0482 & $67.8$ & 0.961& 0.83 \\
APOSTLE (IR) &0.272  & 0.728& 0.0455& $70.4$ & 0.967 & 0.81 \\          
MaGICC & 0.24 & 0.76 & 0.04 & 73 & -- & 0.79\\
Sloane {\it et al.} &0.26& 0.74 & 0.0455 & $73$ & 0.96 & 0.77 \\
\hline
\end{tabular}
\caption{Cosmological parameters used to generate the initial conditions in different simulations.} 
\label{tab:CosmoPar}
\end{table}

\subsection{Ling {\it et al.}}

The simulation performed by Ling {\it et al.}~\cite{Ling:2009eh} uses the cosmological Adaptive Mesh Refinement code {\sc ramses}~\cite{Teyssier:2001cp}, and a ``zoom-in" technique to re-simulate at a higher resolution selected DMO halo. The mass and spatial resolution decreases as the distance from the central region increases. 

In this simulation, a standard Schmidt law is used to implement star formation by generating stars as a Poisson random process~\cite{Rasera:2005gq}. A second-order unsplit Godunov scheme is used to describe gas dynamics, and radiative gas cooling is implemented based on a standard equilibrium photo-ionised mixture of hydrogen and helium, where excess cooling due to metals is taken into account~\cite{Katz:1992db, Sutherland:1993ec}. Supernova feedback is taken into account using the recipe described in Ref.~\cite{Dubois:2008mz}.

\subsection{Eris}

The Eris hydrodynamical simulation~\cite{Kuhlen:2013tra} is a cosmological zoom-in simulation of a single MW analogue. 
The Eris simulation run (details in Ref.~\cite{Guedes:2011ux}) was performed using the N-body$+$smooth particle hydrodynamics (SPH) code {\sc gasoline}~\cite{Wadsley:2003vm} and includes galaxy formation physics. The end product of their simulation is a barred, late-type spiral galaxy similar to the MW. 

The baryonic processes included in the simulation are: star formation based on atomic gas density threshold of 5 atoms cm$^{-3}$ with 10\% star formation efficiency, radiative cooling at low temperatures which is Compton, atomic, and metallicity dependent, heating from supernova explosions and a cosmic ultraviolet field, and a ``blastwave" scheme for supernova feedback~\cite{Stinson:2006cp}. In the blastwave model, the thermal heating of gas from supernovae is mimicked by locally suspending the gas cooling. Notice that gas metal cooling is ignored in the Eris simulation, and this may affect the DM distribution in the Solar neighborhood.

In addition to the Eris hydrodynamical simulation, this study includes a counterpart DMO simulation, ErisDark, which shares the same halo formation history as Eris but treats all matter as collisonless DM.

\subsection{NIHAO}

The NIHAO (Numerical Investigation of a Hundred Astrophysical Objects) simulations~\cite{Wang:2015jpa} are a large suite of cosmological zoom-in simulations, and use an updated version of the MaGICC simulations~\cite{Stinson:2012uh} for including baryonic processes. The initial conditions are generated using the Planck 2014 cosmological parameters as listed in Table~\ref{tab:CosmoPar}. 
An improved version of the SPH code {\sc gasoline}~\cite{Wadsley:2003vm} is used, namely the {\sc ESF-Gasoline2} code which includes a revised treatment of the hydrodynamics~\cite{2014MNRAS.442.3013K}. The same numerical resolution is maintained across the wide mass range in the simulations, from dwarf galaxies to massive MW-like spiral galaxies. The galaxies in the NIHAO simulations reproduce the observed stellar mass-halo mass relation.

The improved version of {\sc gasoline} includes  a subgrid model for turbulent mixing of metals and energy~\cite{Wadsley:2008}, photoelectric heating of dust grains, ultraviolet heating and ionization and cooling due to hydrogen, helium and metals~\cite{Shen:2009zd}. The star formation and feedback is similar to the model used in the MaGICC simulations~\cite{Stinson:2012uh}. Star formation occurs for a gas temperature and density threshold of 15000 K and 10.3 atoms cm$^{-3}$, respectively. Blastwave supernova feedback mechanism~\cite{Stinson:2006cp} is used, and the cooling function is affected by metals which are produced by type II and type Ia supernovae~\cite{Shen:2009zd}. 

Each halo in the NIHAO project is initially simulated at high resolution without baryons, and thus has a DMO counterpart.

Notice that in Table~\ref{tab:sims}, we only list the gas and DM particle masses for one of the MW-like haloes (`g1.92e12') which is the only information presented in Ref.~\cite{Butsky:2015pya}.

\subsection{EAGLE and APOSTLE}

The simulations in the EAGLE project~\cite{Schaye:2015,Crain:2015} were performed using a state-of-the-art hydrodynamical SPH implementation, {\sc anarchy} \cite{DallaVecchia:2015,Schaller:2015b} which was built on top of an optimized version of the SPH code {\sc gadget}~\cite{Springel:2005}, as well as a detailed  subgrid model of galaxy formation. The parameters of the subgrid model are calibrated to produce the observed relation between galaxy stellar mass and size for disk galaxies at $z=0.1$. The Planck 2013~\cite{Planck:2014}  cosmological parameters are assumed (see Table~\ref{tab:CosmoPar}). In Ref.~\cite{Bozorgnia:2016ogo}, the implications for direct detection are studied  using the EAGLE high resolution (HR) simulation which in the EAGLE papers is referred to as Recal-L025N0752.

The APOSTLE project~\cite{Sawala:2015} uses the same code as EAGLE, and applies it to a series of zoomed regions containing DM halo pairs analogous to the MW-M31, or Local Group, system. We use the twelve intermediate resolution APOSTLE volumes, which we refer to as APOSTLE IR, and are comparable in resolution to EAGLE HR\footnote{Higher resolution simulations, denoted as APOSTLE HR also exist within the APOSTLE project, but were not used since their  stellar masses are lower than the observed MW stellar mass range.}. These simulations were run using the WMAP7 cosmological  parameters, given in Table~\ref{tab:CosmoPar}.

The simulations of the EAGLE project use state-of-the-art subgrid models and numerical techniques to include star formation and its energy feedback, radiative cooling, stellar mass loss and metal enrichment, the gas accretion and mergers of supermassive black holes, as well as AGN feedback. Feedback efficiencies are calibrated to reproduce the present day stellar mass function and the observed relation between stellar mass and BH mass, taking into account the galaxy sizes. Feedback from massive stars and AGN are significantly improved compared to previous simulations, such that thermal energy is injected into the gas  without turning off hydrodynamical forces or radiative cooling. As a result, galactic winds are generated without the need to specify wind directions, velocities, or other information. 

For the EAGLE and APOSTLE simulations, companion DMO simulations were run treating all the matter content as collisionless. Therefore we can directly compare the DM distribution of galaxies in the hydrodynamical simulation with their DMO counterparts.

\subsection{MaGICC}

The MaGICC (Making Galaxies in a Cosmological Context) hydrodynamical simulations~\cite{Stinson:2012uh} were carried out using the SPH code {\sc gasoline}~\cite{Wadsley:2003vm}. Initial conditions are generated from WMAP3 cosmology as listed in Table~\ref{tab:CosmoPar}. 
A sample of MW-like haloes are identified from a DMO simulation, and were re-evolved with high resolution from high redshift, including baryonic physics. High resolution particles are added in the region of interest using the zoom-in technique, while other regions contain lower resolution particles.

Low-temperature metal cooling~\cite{Shen:2009zd}, ultraviolet background radiation, and a Schmidt-Kennicutt star formation law~\cite{Kennicutt:1997ng} are included in the simulations. The blastwave model for supernova feedback~\cite{Stinson:2006cp} is implemented, as well as early energy feedback from massive stars into the interstellar medium. The early feedback results in simulated galaxies which have small realistic bulges and do not appear to exhibit any problems stemming from overcooling. 

In addition to the hydrodynamical simulations, one halo in a DMO simulation is also studied in MaGICC. This halo was generated with cosmological initial conditions identical to one of the haloes in the hydrodynamical simulations. 

\subsection{Sloane {\it et al.}}

Sloane {\it et al.}~\cite{Sloane:2016kyi} study galaxies which were simulated using the N-body$+$ SPH code {\sc gasoline}~\cite{Wadsley:2003vm}. Four galaxies are selected and re-simulated using the zoom-in technique. The initial conditions used in the simulation assume the WMAP3 cosmological parameters (in Table~\ref{tab:CosmoPar}).  
The simulations include blastwave supernova feedback~\cite{Stinson:2006cp} and metal line cooling~\cite{Shen:2009zd}.

A DMO version of each galaxy exists which is simulated with the same initial conditions, but does not include baryonic physics.

\begin{table}
\centering
\begin{tabular}{@{}cccccc@{}} 
\hline
Simulation & Count & $M_{\rm star}~[\times 10^{10} \Msun]$ & $M_{\rm halo}~[\times 10^{12} \Msun]$ & $\rho_\chi$ [GeV$/$cm$^3]$ & $v_{\rm peak}$ [km$/$s] \\ 
\hline
Ling {\it et al.} & 1 &  $\sim 8$ & $0.63$ & 0.37--0.39 & 239 \\
Eris & 1 & $3.9$ & $0.78$ & 0.42 & 239 \\
NIHAO& 5 & 15.9 & $\sim 1$ & 0.42 & 192--363 \\
EAGLE (HR) & 12 & 4.65--7.12 & 2.76--14.26 & 0.42--0.73 & 232--289 \\
APOSTLE (IR)  & 2 & 4.48, 4.88 & 1.64--2.15 & 0.41--0.54 & 223--234 \\          
MaGICC & 2 & 2.4--8.3 & 0.584, 1.5 & 0.346, 0.493 & 187, 273 \\
Sloane {\it et al.} & 4 & 2.24--4.56& 0.68--0.91 & 0.3--0.4 & 185--204 \\ 
\hline
\end{tabular}
\caption{The number and properties of the selected MW analogues in the simulations discussed in this paper. The columns specify the number of the MW analogues, (the range of) their stellar mass $M_{\rm star}$, halo mass $M_{\rm halo}$, local DM density $\rho_\chi$, and best fit peak speed of the Maxwellian distribution, $v_{\rm peak}$, for each simulation.}
\label{tab:MW-like}
\end{table}


\section{Identifying simulated Milky Way analogues}
\label{IdentifyMW}

To make precise quantitative predictions for the DM distribution from simulations, one needs to identify simulated galaxies which satisfy MW observational constraints. 
The criteria used to select simulated galaxies which resemble the MW are widely different among different simulation groups, and we review them in this section. The number and properties of the MW analogues selected in each simulation are presented in Table~\ref{tab:MW-like}. Notice that the halo mass has a different definition  in each simulation as described below.

\subsection{Ling {\it et al.}}

The MW analogue in the simulation performed by Ling {\it et al.} has a total mass of $6.3 \times 10^{11}~\Msun$ within a  virial radius defined as the radius containing a density equal to 200 times the critical density. This halo has a steady accretion rate and no major mergers in the last 8 Gyr. The bulge mass is close to the disk mass of $4 \times 10^{10}~\Msun$, which deviates from the MW observational constraints.

\subsection{Eris}

One high resolution simulated MW analogue exists in Eris. At redshift zero this MW analogue has a local circular velocity of 205 km$/$s, a bulge to disk ratio of 0.35, a stellar mass of  $3.9 \times 10^{10}~\Msun$, and a virial mass of $7.8 \times 10^{11}~\Msun$. The virial mass is defined as the mass enclosed within the sphere that contains a mean density of 98 times the critical density of the Universe. 

\subsection{NIHAO}

Four ``test'' galaxies are considered in Ref.~\cite{Butsky:2015pya} with masses between $5 \times 10^{10}$ and $10^{12}~\Msun$ within a virial radius defined as the radius containing a density equal to 200 times the critical density of the Universe. From these test galaxies, one with halo mass of $10^{12}~\Msun$ (called `g1.92e12') is considered as MW-like. Additionally, the DM velocity distribution is presented for four more galaxies (called `g8.26e11', `g1.12e12', `g1.77e12', and `g2.79e12'), which have been chosen to have a halo mass of $\approx 10^{12}~\Msun$ and a stellar mass similar to the MW.

Notice that in Table~\ref{tab:MW-like}, we only list the stellar mass for g1.92e12 which is presented in Ref.~\cite{Butsky:2015pya}. For the local DM density we present in Table~\ref{tab:MW-like} the mean of the average local DM density for all galaxies with halo masses in the range of ($7.5 \times 10^{11} - 3.5 \times 10^{12}$)~$\Msun$ as discussed in  Section~\ref{DMdensity}.

\subsection{EAGLE and APOSTLE}

In the EAGLE HR and APOSTLE IR simulations, a total of 14 MW analogues (called `E1' to `E12' for EAGLE HR and `A1', `A2' for APOSTLE IR in Ref.~\cite{Bozorgnia:2016ogo}) are identified. We first identify all haloes in the mass range $5\times10^{11}<M_{200}/\Msun<2\times10^{13}$, where $M_{200}$ is defined as the mass enclosed within the sphere that contains a mean density 200 times the critical density. We then impose two additional selection criteria for identifying MW analogues: (i) The rotation curves of the simulated galaxies fit well the observed MW rotation curve from Refs.~\cite{Iocco:2015xga}, \cite{Pato:2017yai}. (ii) The stellar mass of the simulated galaxies falls within the 3$\sigma$  MW stellar mass range derived from observations, $4.5 \times10^{10}<M_{*}/\Msun<8.3 \times10^{10}$~\cite{McMillan:2011wd}. 

As discussed in Ref.~\cite{Bozorgnia:2016ogo}, these two criteria are chosen from among many to define a MW analogue, others of which include star formation rate and age, because they directly affect the local circular speed, and therefore the peak of the local DM speed distribution in the Galactic reference frame (see for example Fig.~1 of Ref.~\cite{Bozorgnia:2016ogo}).

Notice that the halo masses of our selected MW analogues are higher than the MW halo mass, $M_{200, {\rm MW}}=1.2^{+0.7}_{-0.4} \times 10^{12}~\Msun$, expected from abundance matching~\cite{Busha:2010sg}. This is probably a result of the slightly too efficient feedback in the simulated haloes in EAGLE HR~\cite{Schaye:2015, Crain:2015}. As discussed in Ref.~\cite{Bozorgnia:2016ogo}, the large halo masses and the mismatch between halo mass and stellar mass do not affect the implications for DM direct detection. In particular, over the small halo mass range probed, we find little correlation between the halo mass and the local DM density or velocity distribution~\cite{Bozorgnia:2016ogo}.

\subsection{MaGICC}

Two simulated MW-like disk galaxies (called `g1536' and `g15784' in Ref.~\cite{Kelso:2016qqj}) are identified with different accretion histories in the MaGICC simulations~\cite{Kelso:2016qqj}. Their virial masses are $5.84 \times 10^{11} \Msun$ and $1.50 \times 10^{12} \Msun$, where the virial radius is defined as the radius containing a density equal to $\sim 100$ times the critical density. Galaxies with masses between $\sim 5 \times 10^{11}$ and $\sim 2 \times 10^{12} \Msun$ were first selected at random from a DMO simulation, requiring that at $z=0$ there was no structure with a mass greater than $\sim 5 \times 10^{11} \Msun$ within 2.7 Mpc of the halo. The simulations for the regions containing the selected galaxies were re-evolved with baryonic physics. 

Both MW-like galaxies have present day halo masses similar to that of the MW. The stellar mass of g1536 is lower than the observed MW stellar mass, whereas g15784 has slightly more stellar mass than the MW. The two galaxies especially differ from one another in their stochastic accretion histories and in the initial conditions from which they evolved. The last merger of g1536 occurred at $z=2.9$, while g15784 had its last major merger at $z=2$. The DMO simulation (called `g1536DM' in Ref.~\cite{Kelso:2016qqj}) was generated with the same initial conditions  as g1536. Notice that the rotation curves of all three haloes differ from the MW rotation curve due to their different total mass distributions~\cite{Kelso:2016qqj}.

\subsection{Sloane {\it et al.}}

Four simulated MW analogues (called `h239', `h258', `h277', and `h285' in Ref.~\cite{Sloane:2016kyi}) with virial masses\footnote{It is not clear how the virial mass is defined in Ref.~\cite{Sloane:2016kyi}.}  in the range of (0.7 -- 0.9) $\times 10^{12} \Msun$ are selected in the simulation performed by Sloane {\it et al.} The selected galaxies are chosen to span a range of merger histories, with three out of four chosen to have recent mergers. Only h277 has a relatively quiescent merger history similar to the MW, with its last merger at $z \sim 3$. h239 was continually bombarded with small galaxies, and h285 underwent recent mergers at $z \sim 1.7$ until $z \sim 0.8$, including one counter-rotational merger. Finally, h258 which has  a prominent dark disk (see the discussion in Section~\ref{VelComponents}), had a same mass merger at $z \sim 1$. This is the same halo studied in Ref.~\cite{Read:2009iv} at a lower resolution compared to Sloane {\it et al.}~\cite{Sloane:2016kyi}.


\section{Local dark matter density and halo shape}
\label{DMdensity}

After having selected the simulated MW analogues, the next step is to extract the DM density in the Solar neighborhood for each simulated halo. To find the DM density at the Solar circle, a region around $\sim 8$ kpc from the Galactic center has to be identified. Each simulation group uses a slightly different region at the Solar circle to determine the average local DM density. Current hydrodynamical simulations are limited in resolution, and cannot resolve the local variations in the DM distribution at different azimuthal angles around the Solar circle. Therefore, the DM density is usually averaged over a cylindrical or spherical region around the Solar circle. 

Studying the shape of the inner DM halo helps our understanding of how baryonic physics affect the local DM density. In most simulations, the sphericity, $s$, of the DM halo which is defined as the ratio of the minor to major axis of the halo is determined. If the DM halo is a perfect sphere, $s=1$, while $s<1$ denotes a departure from a spherical shape. 

\subsection{Ling {\it et al.}}

The average DM density in the Solar neighborhood is found in a torus aligned with the stellar disk with galactocentric radii between 7 and 9 kpc and height between -1 kpc and 1 kpc with respect to the Galactic plane. Alternatively, a torus with the same radii, but a height between -0.1 kpc and 0.1 kpc is considered. The average local DM density is 0.37 and 0.39 GeV$/$cm$^3$, for the larger and smaller torus, respectively. Due to the dark disk component (discussed in Section~\ref{VelComponents}), the shape of the DM halo is oblate.

\subsection{Eris}

The Eris simulation considers a cylindrical volume aligned with the stellar disk with a height extending from -0.1 kpc to 0.1 kpc. The DM density profile is calculated in evenly spaced logarithmic bins in the cylinder. The DM density in the cylinder at 8 kpc is 0.42 GeV$/$cm$^3$, which is higher by 34\% than the spherically average DM density at 8 kpc, and higher by 31\% compared to the spherically averaged DM density in the ErisDark DMO simulation. This is a result of both a contraction of the DM halo due to dissipational baryonic processes occurring in the plane of the Galactic disk, as well as the natural tendency of galaxies to be aligned with the halo in which they live. The DM halo is oblate with intermediate to minor axis ratio of 0.99 and minor to major axis ratio of $s=0.69$, which is rounder and more axisymmetric compared to the DM halo in ErisDark. 

\subsection{NIHAO}

The DM density profile is presented individually for the four test galaxies specified in Ref.~\cite{Butsky:2015pya}. For g1.92e12 which has a mass similar to the MW, the spherically averaged DM density is 0.7 GeV$/$cm$^3$ at 8 kpc, as read off from Fig.~11 of Ref.~\cite{Butsky:2015pya}. This is similar to the local DM density of the DMO counterpart of this halo, as seen from the same figure. The average DM density for all galaxies in the halo mass range of $7.5 \times 10^{11}<M_{\rm halo}<3.5 \times 10^{12}~\Msun$ is presented in Fig.~12 of Ref.~\cite{Butsky:2015pya}. At 8 kpc, the mean value of the average DM density is 0.42 GeV$/$cm$^3$, as read off from the top right panel of that figure.

The shape of the inner halo measured at 12\% of the virial radius, for the MW-like galaxies (mass of $\approx 10^{12}~\Msun$) in the hydrodynamical simulation is close to spherical, and has an average minor to major axis ratio of $s=0.8$. This is rounder than the shape of the inner DM halo in the DMO simulation which is triaxial. As expected, the effect of the baryonic processes which modify the shape of the DM distribution is strongest in the inner halo.

\subsection{EAGLE and APOSTLE}

We find the local DM density of the simulated MW analogues in the EAGLE HR and APOSTLE IR simulations, by considering a torus aligned with the stellar disk, with an inner and outer radii of 7 and 9 kpc from the Galactic center, respectively, and a height from -1 to 1 kpc with respect to the Galactic plane. The average local DM density in this torus is 0.42 -- 0.73 GeV$/$cm$^3$ for the 12 haloes in the EAGLE HR simulation, and 0.41 -- 0.54 GeV$/$cm$^3$ for the two haloes in APOSTLE IR. The average DM density varies on average by 32\% along the torus, which is smaller than the halo-to-halo scatter. For two out of 14 MW analogues, the local DM density in the torus is greater than 20\% compared to the average local DM density in a spherical shell with radius between 7 and 9 kpc. Such an increase could be a result of the DM halo contraction due to dissipational baryonic physics. 

We find that at 5 kpc, the haloes are close to spherical with sphericity, $s=[0.85,0.95]$. The sphericities are lower by less than 10\% at 8 kpc. The deviation of the halo shapes towards either prolate or oblate distributions is very small. The DM sphericity is higher in the hydrodynamical simulations compared to the DMO case, in agreement with the result of earlier simulations~\cite{Dubinski:1993df, Bryan:2012mw}, and higher resolution APOSTLE simulations~\cite{Schaller:2015mua}.

\subsection{MaGICC}

To find the local DM density for the two MW-like galaxies in the MaGICC simulations, a torus at the Solar circle is considered with a major radius of 8 kpc and a minor radius of 2 kpc. The average DM density in the torus is 0.346 and 0.493 GeV$/$cm$^3$ for the two galaxies.

The shape of one of the DM haloes (g1536) in MaGICC hydrodynamical simulations is close to axisymmetric with an intermediate to major axis ratio of $\sim 1$, and flattened with an intermediate to minor axis ratio of $\sim 0.75$. This halo has a sphericity of $s \sim 0.75$ within 5 kpc, as read from the middle panel of Fig.~1 in Ref.~\cite{Kelso:2016qqj}. The other halo (g15784) is nearly spherical with $s \sim 0.9$  within 50 kpc. The differences in the shape of the two haloes are a result of their different accretion histories~\cite{Kelso:2016qqj}. The shape of the halo in the DMO simulation (g1536DM) is highly prolate, except in the inner 2 kpc region. Due to the absence of baryons, the DM halo in the DMO simulation is also triaxial.

\subsection{Sloane {\it et al.}}

Sloane {\it et al.} consider a cylindrical annulus in the plane of the stellar disk with a central radius of 8 kpc, and height and width of 1 kpc. The average local DM density is $\sim 0.3$ -- 0.4 GeV$/$cm$^3$ for the haloes they consider\footnote{It is not clear if this is the average DM density in the cylindrical annulus region or in a spherical shell around the Solar circle.}. The shape of the DM haloes are not discussed in Ref.~\cite{Sloane:2016kyi}.


\section{Local dark matter velocity distribution}
\label{f(v)}

In order to determine if the SHM provides an accurate description for the DM halo in the Solar neighborhood, one can compare the DM velocity distribution of the simulated MW analogues in the Galactic rest frame to a Maxwellian velocity distribution with a peak speed of 220 km/s, which is usually assumed in the direct detection community. In case of deviations from the SHM, the DM velocity distribution can be fitted with a Maxwellian distribution with a free peak speed, as well as with various other fitting functions.

The velocity vector of the simulation particles is specified in a reference frame in the plane of the galaxy, with origin at the Galactic center, $r$-axis in the radial direction, $\theta$ in the tangential direction, and the $z$-axis perpendicular to the stellar disk. The DM velocity modulus (speed) distribution, as well as the distributions of the radial ($v_r$), azimuthal ($v_\theta$), and vertical ($v_z$) components of the DM velocity distribution can be calculated. The DM speed distribution and the three components of the velocity distribution are individually normalized to unity, such that $\int dv f(v)=1$ and $\int dv_i f(v_i)=1$ for $i=r, \theta, z$. Notice that the speed distribution, $f(v)$, is related to the velocity distribution, $f(\bf v)$, by $f(v) = v^2 \int d\Omega_{\bf v} f(\bf v)$, where $d\Omega_{\bf v}$ is an infinitesimal solid angle around the direction of the DM velocity  ${\bf v}$.


\subsection{DM speed distribution}
\label{SpeedDist}

\begin{figure}[t]
\includegraphics[width=0.49\textwidth]{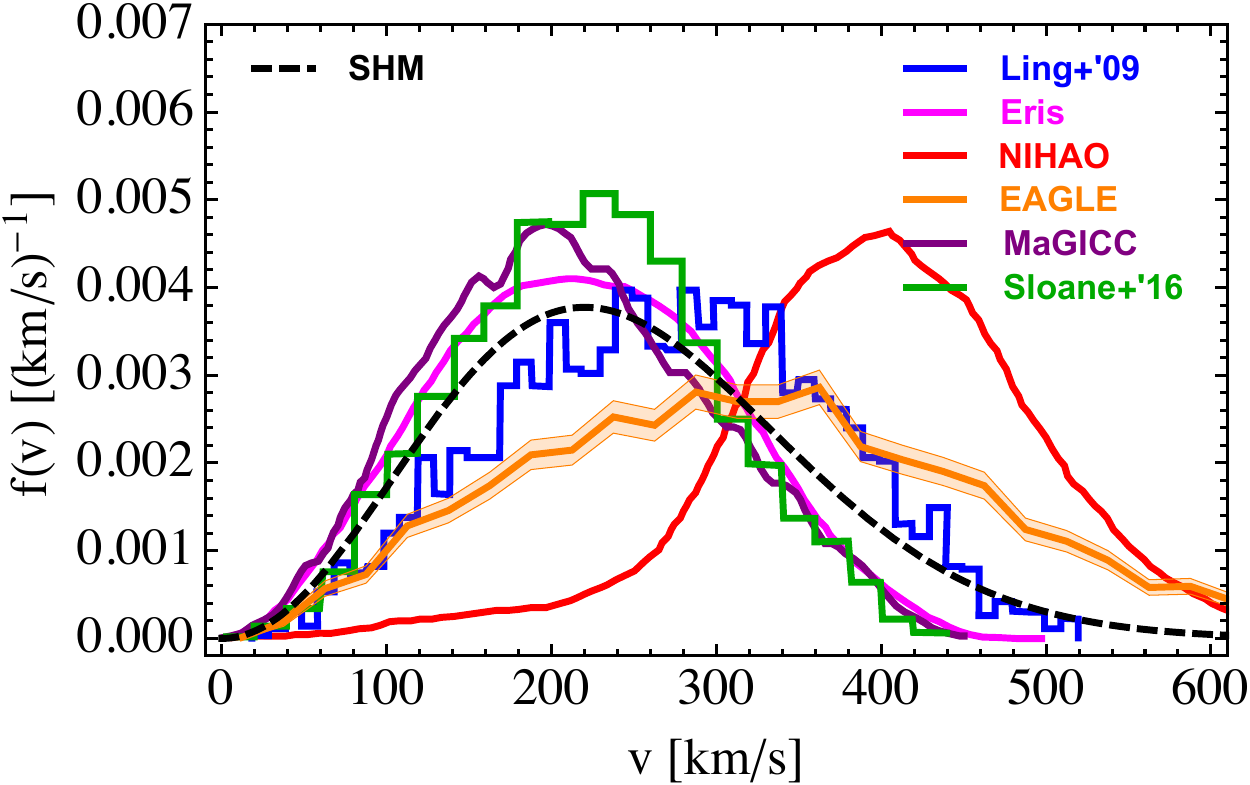}
\includegraphics[width=0.49\textwidth]{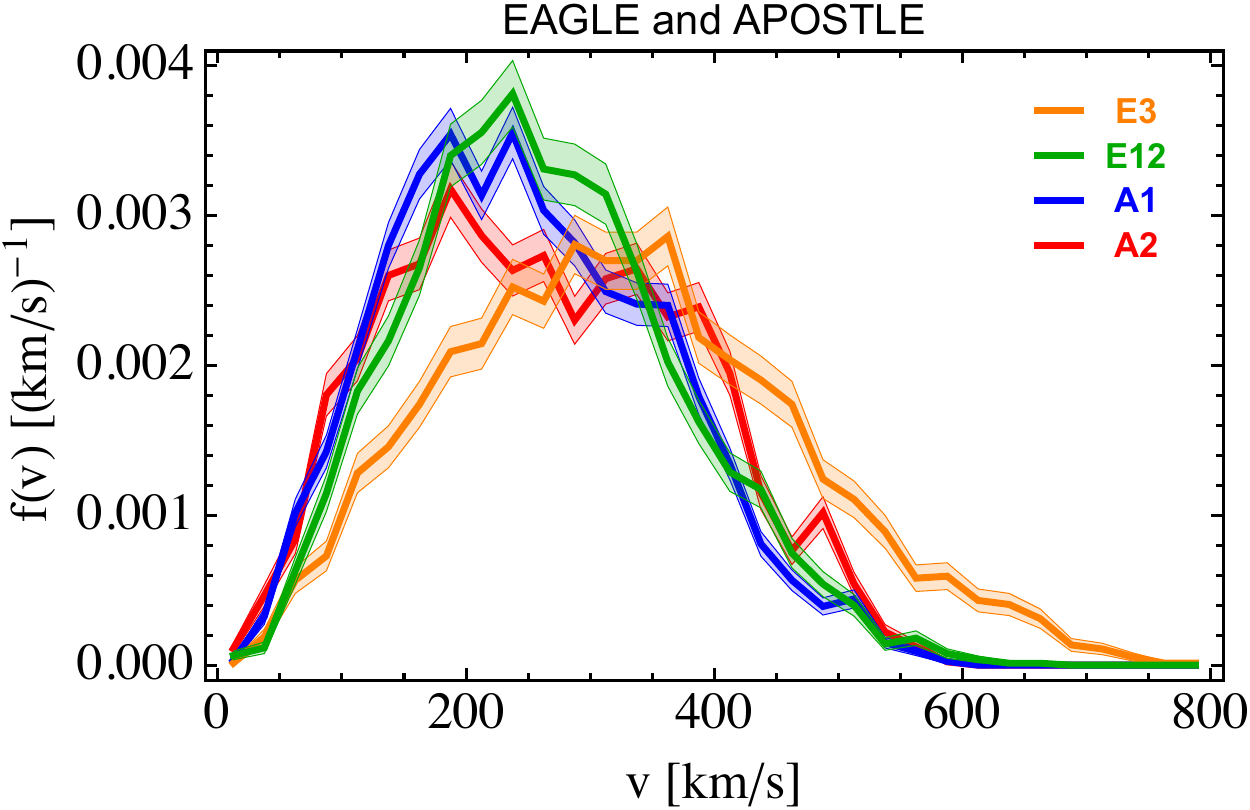}
\caption{Local DM speed distributions in the Galactic rest frame for (a) MW-like haloes in different hydrodynamical simulations (solid colored lines) which have the farthest speed distribution from the SHM Maxwellian with a peak speed of 220 km$/$s (dashed black line), and (b) two MW-like haloes in the EAGLE HR simulation which have speed distributions closest to (green) and farthest from (orange) the SHM and two haloes in the APOSTLE IR simulation (blue and red). The solid curves and shaded regions specify the mean of the DM speed distribution and its 1$\sigma$ error, respectively. The speed distributions plotted on the left panel are taken from Refs.~\cite{Ling:2009eh, Kuhlen:2013tra, Butsky:2015pya, Bozorgnia:2016ogo, Kelso:2016qqj, Sloane:2016kyi}.
\label{fig:fv}}
\end{figure}

As expected, due to the differences in hydrodynamical approach, cosmological parameters, and definitions of physical quantities, the simulations we analyzed exhibit a variety of local speed distributions. We found in particular that:

\begin{itemize}

\item {\bf There exists a large variation in local speed distributions between the results of different simulations}. To illustrate this point, we show in the left panel of Fig.~\ref{fig:fv} the local speed distributions in the Galactic reference frame  for one MW-like halo from each hydrodynamical simulation which has the farthest speed distribution from the SHM Maxwellian with a peak speed of 220 km$/$s (shown as a black dashed curve in the same plot). These haloes are g2.79e12 from NIHAO, E3 from EAGLE HR, g1536 from MaGICC, and h258 from Sloane {\it et al}. 


\item {\bf There is a substantial halo-to-halo variation in local 
DM speed distributions within a given simulation suite}. We demonstrate this point in the right panel of Fig.~\ref{fig:fv}, where we present the local DM distributions of two haloes in the EAGLE HR simulations which have speed distributions closest to (halo E12, shown in green) and farthest from (halo E3, shown in orange) the SHM, as well as the two haloes in the APOSTLE IR simulations (haloes A1 and A2, shown in blue and red, respectively). The shaded bands specify the 1$\sigma$ uncertainty in the speed distribution from each halo.

\end{itemize}

A number of fitting formulae have been proposed in the literature to parametrize the departure from the standard Maxwellian, including:

\begin{itemize}

\item Generalized Maxwellian distribution~\cite{Ling:2009eh}:
\beq \label{eq:genMax}
f(v) \propto  v^2  \exp[ -( v / v_0 )^{2 \alpha} ] \, , 
\eeq
with free parameters $v_0$ and $\alpha$. A standard Maxwellian
distribution is recovered for $\alpha = 1$. 

\vspace{15pt}

\item The speed distribution proposed by Mao {\it et al.}~\cite{Mao:2012hf}:
\beq \label{eq:Mao}
f(v) \propto v^2  \exp[ - v/ v_0 ]  ( v_{\rm esc}^2 -  v^2)^p ~\Theta (v_{\rm esc} - v)\, , 
\eeq
with free parameters $v_0$ and $p$.

\vspace{15pt}

\item The Tsallis~\cite{Tsallis:1987eu} distribution:
\beq \label{eq:Tsallis}
f(v) \propto v^2  \left(1 - (1-q) v^2/v_0^2 \right)^{q/(1-q)}\, , 
\eeq
with free parameters $v_0$ and $q$. We obtain the standard Maxwellian
distribution for $q \rightarrow 1$.

\end{itemize}

Notice that all the fitting functions are normalized such that $\int_0^{v_{\rm esc}} dv f(v)=1$.

Below, we describe how well the local DM speed distributions of the MW analogues in each simulation fit a Maxwellian distribution with a free peak as well as the above fitting formulae.  

\subsubsection{Ling {\it et al.}}

The DM speed distribution for the MW analogue is considered in a torus aligned with the stellar disk with radii between 7 kpc to 9 kpc and a height between -1 kpc and 1 kpc. There are 2662 particles in this region. Strong deviations from the best fit Maxwellian speed distribution ($\alpha =1$ in Eq.~\ref{eq:genMax}) is observed, with a large deficit at large speeds. Although a Tsallis distribution (Eq.~\ref{eq:Tsallis}) provides an  excellent fit to the DM particles in a spherical shell around 8 kpc, it does not provide a good fit to the particles in the torus around the Solar position.

\subsubsection{Eris}

To find the average DM speed distribution in the Galactic rest frame at the Solar position, Eris considers an annulus aligned with the stellar disk with inner and
outer radii of 6 kpc and 10 kpc from the Galactic center, respectively, and with a height spanning from -2 kpc to 2 kpc. There are 81,213 DM particles in this region. The DM speed distribution in the annulus has a mean speed of 220.8 km$/$s, and is compared to the SHM consisting of a Maxwellian velocity distribution with a peak  speed of 220 km$/$s and a Maxwellian velocity distribution with the same peak speed as the simulation. At all speeds less than 350 km$/$s, the DM speed distribution in the simulation is larger than the SHM, and decreases sharply at higher speeds. The DM speed distribution of the simulation is not a perfect fit to the matched to peak Maxwellian distribution, with a slight excess at 230 to 380 km$/$s, and a deficit at higher speeds. The distribution instead fits well the fitting function proposed by Mao  {\it et al.} (Eq.~\ref{eq:Mao}).

To identify the effect of baryons on the local DM distribution, a comparison with the ErisDark DMO simulation is performed in Ref.~\cite{Kuhlen:2013tra}. In ErisDark, the average DM speed distribution is found in a spherical shell of width 4 kpc around the Solar circle. The DM speed distribution shows the usual departures from the Maxwellian distribution seen in DMO simulations, i.e.~less particles close to the peak and an excess of particles at high speeds. The speed distribution is well fit by the Mao {\it et al.} fitting function. Compared to Eris, the DM speed distribution is broadened and shifted to higher speeds. Baryons fall in the center of the DM halo and deepen the Galactic potential well, resulting in more high speed DM particles at the Solar position. The Maxwellian matched to peak is a much better fit to the local DM speed distribution in Eris than in ErisDark, indicating that {\it including baryons in the simulation results in DM speed distributions closer to a Maxwellian functional form}.

\subsubsection{NIHAO}

To find the local DM speed distribution in the Galactic rest frame, the NIHAO collaboration considers a spherical shell with radius between 7 kpc and 9 kpc. The speed distributions from the simulated galaxies are compared to their best fit Maxwellian and Gaussian distributions. They find that compared to the best fit Maxwellian, the speed distributions of the simulated galaxies fall faster at high velocities, and are more symmetric around the peak speed. However, the speed distributions are fitted well by a Gaussian distribution.

The local DM speed distributions for the haloes in the DMO simulations can be well fitted with a Maxwellian distribution. The large difference between the speed distributions in the hydrodynamical and DMO simulations may be due to the strong feedback model adopted in the NIHAO simulations which results in a strong baryonic impact. 
Due to the halo contraction in the hydrodynamical simulations, the peak of the DM velocity distribution moves to higher speeds compared to the DMO simulations.

\subsubsection{EAGLE and APOSTLE}

For each simulated MW analogue in the EAGLE HR and APOSTLE IR simulations, the average DM speed distribution is found in the same torus as described in Section~\ref{DMdensity} used for finding the local DM density. The torus contains a total of 1821 -- 3201 particles depending on the halo in the EAGLE and APOSTLE simulations. The local speed distributions of the MW analogues are fitted with various fitting functions. The Maxwellian distribution with a free peak ($\alpha =1$ in Eq.~\ref{eq:genMax}) provides a better fit to haloes in the hydrodynamical simulations compared to their DMO counterparts. The range of the best fit peak speeds of the Maxwellian distribution is 223 -- 289 km$/$s. The fitting function of Mao {\it et al.}  (Eq.~\ref{eq:Mao}) which has one extra free parameter provides the best fit for almost all MW analogues. 

\begin{figure}[t]
\centerline{\includegraphics[width=0.6\textwidth]{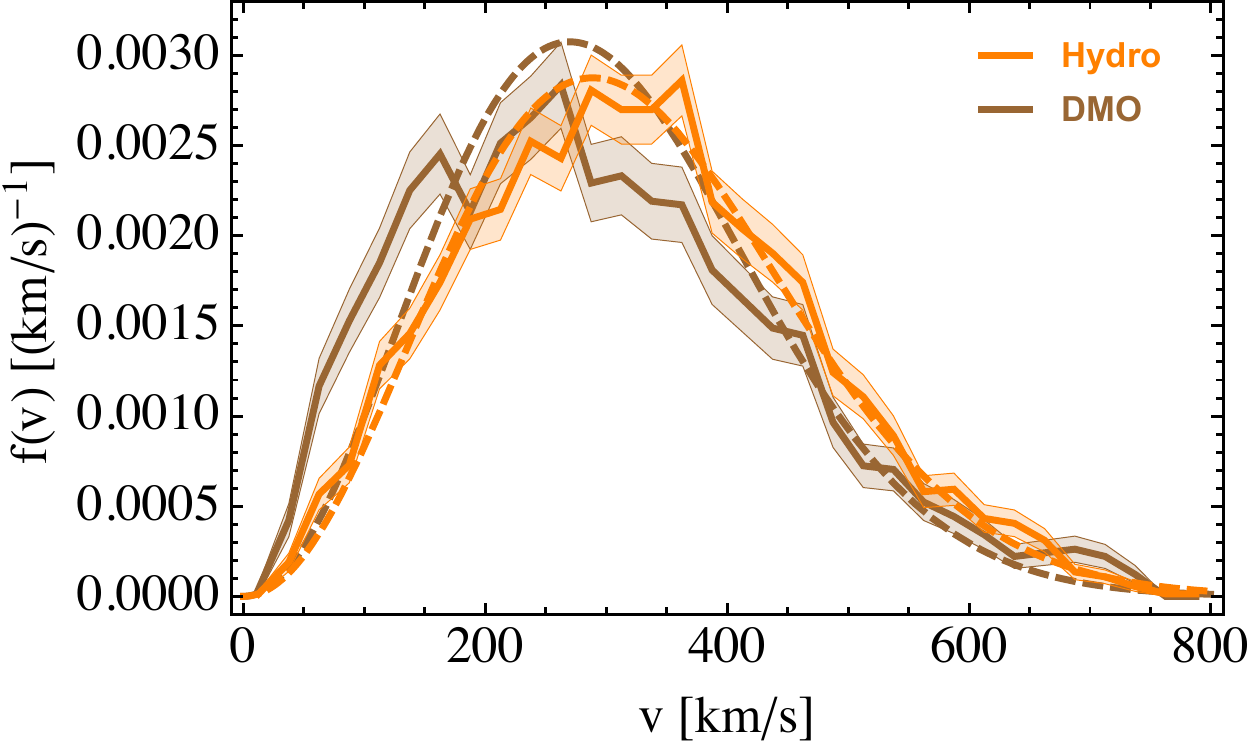}}
\caption{Comparison of the local DM speed distributions in the Galactic rest frame for a MW analogue (halo E3) in the EAGLE HR hydrodynamical simulation (solid orange line and its 1$\sigma$ error band) and its DMO counterpart (solid brown line and its 1$\sigma$ error band). Dashed lines specify the best fit Maxwellian speed distributions for the hydrodynamical (orange) and DMO (brown) case.
\label{fig:HydroDMO}}
\end{figure}

Figure~\ref{fig:HydroDMO} shows the local DM speed distribution for a MW-like halo in the EAGLE hydrodynamical simulation (solid orange line and its 1$\sigma$ error band) which has a speed distribution farthest from the SHM (halo E3) and its best fit Maxwellian speed distribution (dashed orange line). For comparison, the DM speed distribution for the same halo in the DMO simulation (solid brown line and its 1$\sigma$ error band)  and its best fit Maxwellian speed distribution (dashed brown line) is also shown. The best fit Maxwellian speed distribution gives a good fit to the speed distribution of the simulated halo in the hydrodynamical case, but cannot provide a good fit in the DMO case. For most haloes in the EAGLE and APOSTLE DMO simulations, there are large deficits of DM particles at the peak, and an excess at low and very high speeds compared to the best fit Maxwellian distribution. These features are similar to those seen in other DMO simulations~\cite{Vogelsberger:2008qb, Kuhlen:2009vh}.

\subsubsection{MaGICC}

The same torus described in Section~\ref{DMdensity} is considered to find the average DM speed distribution. The region of the torus contains 4849 and 6541 particles for the two MW-like galaxies in the MaGICC simulations. The speed distributions of the simulated haloes are compared to the SHM speed distribution inferred for each halo from its mass distribution. Other than the inferred SHM, two additional approximations are considered: the best fit Maxwellian speed distribution assuming a stationary halo, as well as allowing for a  bulk rotation of the halo. The SHM provides a reasonable fit to the speed distributions of the simulated haloes, except in the high speed tail where there is some deficit of particles from the simulations compared to the SHM. The best fit Maxwellian distributions with and without rotation also provide good fits to the speed distributions from simulations and are almost identical. 

In agreement with previous DMO simulations, the SHM is not a good fit to the speed distribution of the halo in the MaGICC DMO simulation.

\subsubsection{Sloane {\it et. al.}}

Sloane {\it et al.} considers the same cylindrical annulus described in Section~\ref{DMdensity}  to find the local DM speed distribution. There are a total of 5847 -- 7460 DM particles in the annulus, depending on the halo. The average DM speed distribution in this annulus is found for each halo, and compared to the SHM assuming a Maxwellian speed distribution with a peak speed of 220 km$/$s, the best fit Maxwellian speed distribution, and the empirical speed distribution from Mao {\it et al.} (Eq.~\ref{eq:Mao}). Compared to the SHM, there is a deficit of high velocity DM particles in the simulations. However, the SHM is a better fit to the haloes in the hydrodynamical  simulations compared to their DMO counterparts. The best fit Mao {\it et al.} fitting function  provides a better fit to the speed distributions from the simulations, compared to the SHM. 

As expected, including baryons in the simulation increases the gravitational potential in the galactic disk and results in more high speed DM particles compared to the DMO case.


\subsection{Best fit peak speed and local circular speed}
\label{vpeakvc}

\begin{figure}[t]
\centerline{\includegraphics[width=0.6\textwidth]{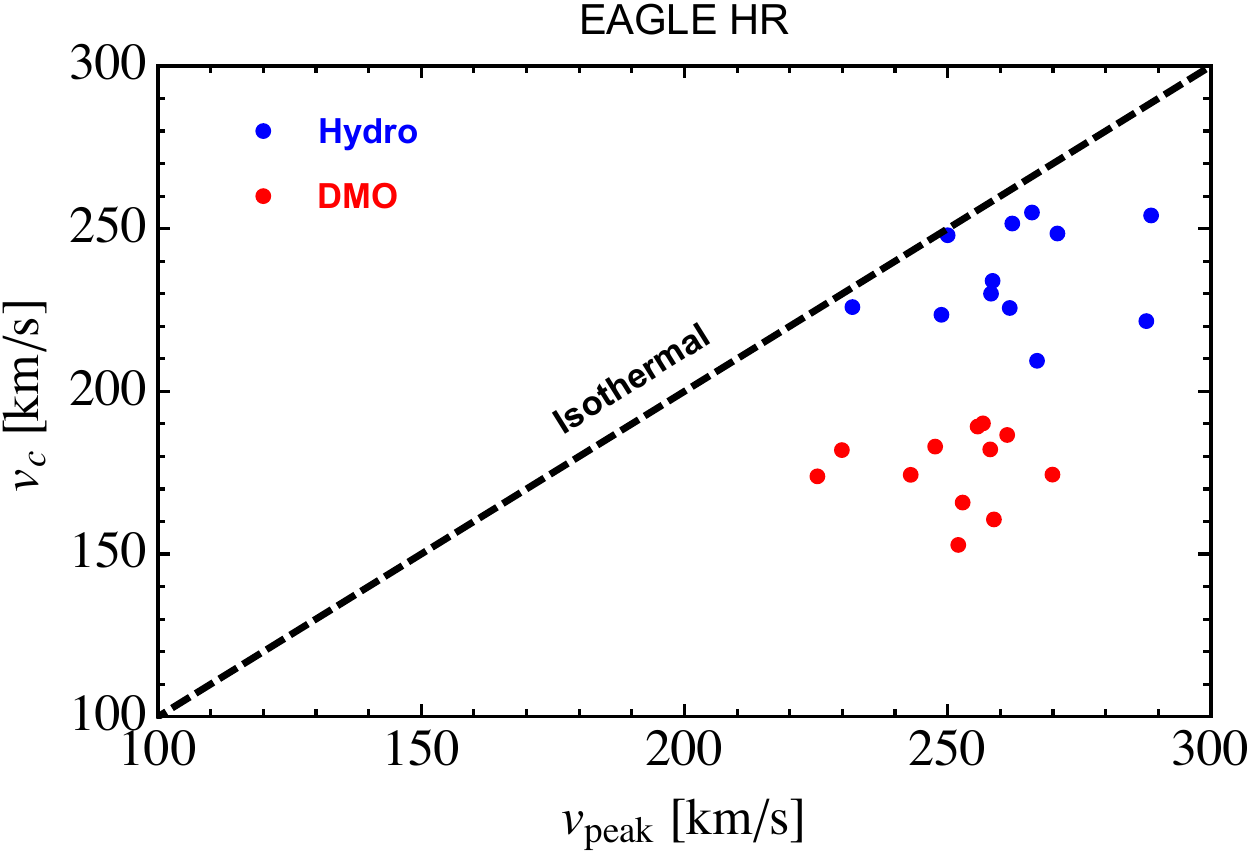}}
\caption{The correlation of the local circular speed, $v_c$, at 8 kpc with the peak speed, $v_{\rm peak}$, of the best fit Maxwellian speed distribution for the MW-like haloes in the EAGLE HR hydrodynamic simulation (blue) and their DMO counterparts (red). The case of an isothermal halo where $v_{\rm peak}=v_c$ is shown as a dashed black line.
\label{fig:vpeakvc}}
\end{figure}

In an isothermal halo, the DM velocity distribution in the Galactic rest frame is a Maxwellian distribution with a peak speed equal to the local circular speed. In Fig.~\ref{fig:vpeakvc} we show how the best fit peak speed, $v_{\rm peak}$, of the Maxwellian speed distribution in the Galactic rest frame for each simulated MW analogue in the EAGLE HR simulation is correlated with the local circular speed, $v_c$, at 8 kpc for that halo. Notice that $v_c$ is computed from the total mass (in stars, gas, and DM) of the halo enclosed within a sphere of radius 8 kpc. For comparison, we also show the correlation of $v_{\rm peak}$ and $v_c$ for the DMO counterparts of the same haloes. 

One can draw a few important conclusions from Fig.~\ref{fig:vpeakvc}. The local circular speeds for all MW analogues in the hydrodynamical simulation are larger than the local circular speed of their DMO counterparts. This is not surprising since the effect of baryons is especially important in the Solar circle and results in an increased $v_c$. As discussed in Ref.~\cite{Bozorgnia:2016ogo}, the enclosed stellar mass within the Solar circle is a large fraction of the total stellar mass, and hence $v_c$ at 8 kpc is strongly correlated with the total stellar mass. The presence of baryons results in deepening the gravitational potential as well as a compression of the DM distribution, leading to a larger $v_c$. Moreover, in the hydrodynamical haloes the rotation curves at 8 kpc are close to reaching their maximum values, whereas the DMO rotation curves are still rising at 8 kpc. Hence, it is also not surprising that at the Solar circle, the MW-like haloes in the hydrodynamical simulation are closer to having an isothermal distribution compared to their DMO counterparts. This effect was also noticed and mentioned in Ref.~\cite{Kelso:2016qqj} for the haloes in the MaGICC simulations. 

Additionally, most MW analogues in the hydrodynamical simulation have a best fit Maxwellian peak speed, $v_{\rm peak}$, larger than their DMO counterparts. As discussed in Section \ref{SpeedDist}, this effect which is also seen in all other simulations discussed in this work (except for Ling {\it et al.} which does not have a DMO counterpart) is also due to baryons deepening the total gravitational potential of the galaxy and resulting in more high speed particles at the Solar circle.

It is clear from Fig.~\ref{fig:vpeakvc}, that assuming an isothermal distribution for the DMO haloes, namely setting the peak speed of the Maxwellian speed distribution equal to the local circular speed inferred from the DMO simulation leads to local DM speed distributions which significantly differ from the true DM distribution of the simulated halo. We demonstrate how this assumption can affect direct detection implications in Fig.~\ref{fig:etaDMOHydro}.

\subsection{DM velocity distribution components and dark disks}
\label{VelComponents}

To assess whether any velocity anisotropy is present at the Solar radius, one can study the radial, azimuthal, and vertical components of the DM velocity distribution in the Galactic frame. In this section we describe how well the DM velocity components of the MW analogues in each simulation fit different fitting functions. In particular, the following fitting functions are considered:

\begin{itemize} 

\item Gaussian function:
\beq\label{eq:Gauss}
f(v_i) = \frac{1}{\sqrt{\pi} v_0 } \exp \left[-(v_i - \mu)^2/v_0^2 \right] \, , 
\eeq
with free parameters $v_0$ and $\mu$.

\vspace{15pt}

\item Generalized Gaussian function:
\beq\label{eq:GenGauss}
f(v_i) = \frac{1}{2 v_0 \Gamma(1 + 1/(2 \alpha))} \exp \left[ -\left((v_i - \mu)^2/v_0^2 \right)^\alpha \right] \, , 
\eeq
with free parameters $v_0$, $\mu$, and $\alpha$. 

\end{itemize}

The functions are normalized such that $\int_{-\infty}^{\infty} dv_i f(v_i)=1$.
 
In cases where there is an asymmetry in the azimuthal component of the DM velocity distribution, it can be fitted with a
double Gaussian function: 
\beq\label{eq:DoubleGauss}
f(v_\theta) = c_1 f_{\rm 1}^{\rm Gauss}(v_\theta; v_1, \mu_1) + c_2 f_{\rm 2}^{\rm Gauss}(v_\theta; v_2, \mu_2) \, , 
\eeq 
with free parameters
$c_1, v_1, \mu_1, v_2$, and $\mu_2$; $c_2$ is instead constrained by requiring
$f(v_\theta)$ to be normalised to 1, $c_1 + c_2 = 1$.

By studying the azimuthal component of the DM velocity distribution in different simulated haloes, we can also search for the existence of a ``dark disk". A dark disk can form when merging satellite galaxies are disrupted by the baryons in the galactic disk. These satellites are dragged towards the galactic plane and disrupted by tidal forces, and their accreted material forms a DM disk co-rotating with the stellar disk~\cite{Read:2008fh}.

The existence of a dark disk can modify the signals in direct detection experiments~\cite{Read:2009iv, Purcell:2009yp, Bruch:2008rx}. In particular if a large fraction of DM particles in the Solar neighborhood are in the disk, the direct detection event rates could be enhanced, especially in the low recoil energy range. Depending on the rotation speed of the DM disk compared to the stellar disk, the phase of the annual modulation signal could also be shifted. In this section we will also review the predictions of recent hydrodynamical simulations  for the existence of a dark disk component in simulated MW-like haloes.

\subsubsection{Ling {\it et al.}}

The  components of the local DM velocity distribution for the haloes studied in Ling {\it et al.} exhibit anisotropy, and a strong deviation from a Gaussian velocity distribution (Eq.~\ref{eq:Gauss}). The mean of the radial and vertical velocity distributions is compatible with zero, while the DM particles exhibit rotation in the azimuthal direction 
with a minimum lag velocity of 75 km$/$s compared to the galactic disk, suggesting the existence of a dark disk component. The tangential velocity distribution is fitted well with a double Gaussian (Eq.~\ref{eq:DoubleGauss}), and the rotating DM component in the disk constitutes $\sim 25$\% of the total local DM density. The small fraction of the dark disk in this halo suggests a quiescent merger history, and is not expected to affect direct detection signals significantly~\cite{Ling:2009eh}.

\subsubsection{Eris}

For the simulated halo in Eris, the radial and vertical velocity distributions at the Solar circle have a mean compatible with zero, while the tangential velocity distribution is skewed towards positive tangential velocities. This is indicative of a population of DM particles co-rotating with the MW disk.  To assess if this co-rotation is due to the presence of a dark disk component, the amount of material accreted from satellite galaxies, deposited into the disk, and co-rotating with the stellar disk is computed. Computed in this way, 9.1\% of the DM density in the disk is due to the dark disk component. If the additional criterion that the rotation speed of the dark disk lies within 50 km$/$s of the stellar rotation speed is applied, the fraction of the dark disk will be reduced to 3.2\%.

The radial and azimuthal DM velocity distributions are narrower in ErisDark compared to Eris in which dissipational baryonic processes have broadened the distributions. However, the vertical velocity distribution in ErisDark is slightly broader than Eris. The asymmetry seen in the azimuthal velocity distribution in Eris does not occur in ErisDark, and the three components of the velocity distribution have a mean compatible with zero.

\subsubsection{NIHAO}
The components of the DM velocity distribution were not analyzed in Ref.~\cite{Butsky:2015pya}. The DM velocity distribution was however found to be the same in a spherical shell of radius between 7 and 9 kpc, and in a torus aligned with the plane of the disk, suggesting that the DM distribution is spherically symmetric. Based on this observation, Ref.~\cite{Butsky:2015pya} concludes that their simulated haloes do not have dark disks.

\subsubsection{EAGLE and APOSTLE}

The three components of the DM velocity distribution for the MW analogues in the EAGLE HR and APOSTLE IR simulations show clear velocity anisotropy. The radial velocity distribution is broader than the vertical and tangential distributions. The distribution of the radial and vertical velocity components are generally peaked around zero, and well fitted with a generalized Gaussian (Eq.~\ref{eq:GenGauss}) with $\alpha \sim 1$ (close to a Gaussian function). The tangential velocity distribution is well fitted with either a Gaussian or generalized Gaussian with $0.6< \alpha < 1$. In five haloes, there is a significant non-zero mean azimuthal speed ($|\mu| > 20$~km$/$s).

For the DMO counterparts of the MW analogues, the distributions of the radial and vertical velocity components peak around zero. The mean of the best fit generalized Gaussian for the azimuthal velocity component in the hydrodynamical case is larger than the corresponding mean in the DMO simulation by 3$\sigma$.

To study if the asymmetry in the azimuthal component of the DM velocity distribution in some haloes in the hydrodynamical case is due to the existence of a dark disk, we compare the azimuthal velocity distribution of the star particles in the torus for each halo with that of DM particles. Only two haloes have a mean azimuthal velocity comparable (within 50 km$/$s) to that of the stars. The azimuthal component of the DM velocity distribution can also be fitted with a double Gaussian (Eq.~\ref{eq:DoubleGauss}). For the two haloes rotating as fast as the stars, the fraction of the rotating component extracted from the double Gaussian is 32\% and 43\%. In summary, from the 14 MW analogues in the EAGLE HR and APOSTLE IR hydrodynamical simulations, there is a hint for the existence of a co-rotating dark disk only for two haloes.

Dark disks have also been extensively searched for in 24 haloes in the APOSTLE IR simulations~\cite{Schaller:2016uot}, two of which are the  MW analogues (based on our criteria defined in Section~\ref{IdentifyMW}) which we have studied in Ref.~\cite{Bozorgnia:2016ogo} and reviewed in this work. Ref.~\cite{Schaller:2016uot} finds that only one out of 24 haloes in APOSTLE IR show evidence for a dark disk. The dark disk component developed for this one halo because of a recent impact with a large satellite, and from MW kinematical data we do not expect such an encounter for our Galaxy.

\subsubsection{MaGICC}

In the two MaGICC MW-like galaxies studied in Ref.~\cite{Kelso:2016qqj}, the SHM inferred from the mass of each simulated halo, as well as the best fit Maxwellian velocity distributions with and without rotation are almost indistinguishable and provide good fits to the components of the velocity distributions of the simulated haloes.

In the DMO simulation, the SHM is a bad fit to all three components of the DM velocity distribution at all velocities. Compared to the SHM, the best fit Maxwellian with and without rotation provide better fits to the DMO distributions, but the discrepancies are still large.

The best fit rotational speed of the DM particles in the torus is at most $\sim 20$~km$/$s, suggesting that the two simulated galaxies do not have a rotating dark disk component. Similar to the conclusion reached in Refs.~\cite{Bozorgnia:2016ogo} and \cite{Schaller:2016uot} for the EAGLE HR and APOSTLE IR simulations, Ref.~\cite{Kelso:2016qqj} concludes that dark disks are not a generic prediction of hydrodynamical simulations.

\subsubsection{Sloane {\it et. al.}}

The components of the DM velocity distribution are not presented in Sloane {\it et al.}~\cite{Sloane:2016kyi}. It is however mentioned that the DM component of some of the galaxies studied in the hydrodynamical simulation show evidence of co-rotation with the stellar disk, pointing to the presence of a dark disk. As discussed in Section~\ref{IdentifyMW}, one of the haloes (h258) had a large recent merger. For one of the other studied haloes (h285), the DM shows evidence of counter-rotation with respect to the stellar disk which is not surprising since the halo had a recent counter-rotational merger. Only one of the selected haloes has a relatively quiescent merger history similar to the MW. 

The DM velocity distributions for the haloes in the DMO simulation are largely isotropic.


\section{Implications for direct detection}
\label{Implications}

In this section we discuss the halo integrals (given in Eq.~\ref{eq:eta}) and direct detection exclusion limits or allowed regions in the plane of DM mass and scattering cross section, as extracted from the Eris, EAGLE and APOSTLE, MaGICC, and Sloane {\it et al.} simulations. Notice that Ling {\it et al.}~\cite{Ling:2009eh} and Butsky {\it et al.}~\cite{Butsky:2015pya} (NIHAO simulations) do not provide plots of halo integrals for their simulated haloes, and are therefore not included in the discussion of this section. 


\subsection{Halo integrals}

\begin{figure}[t]
\includegraphics[width=0.49\textwidth]{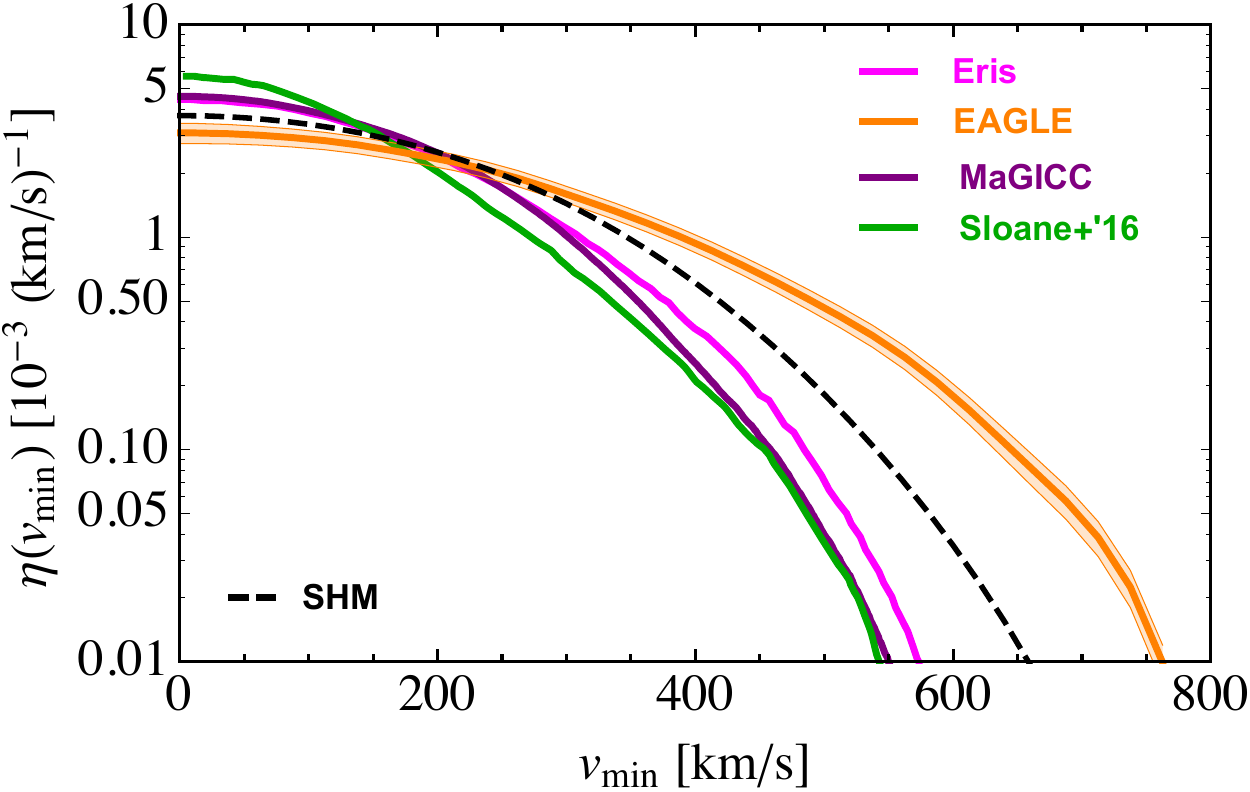}
\includegraphics[width=0.49\textwidth]{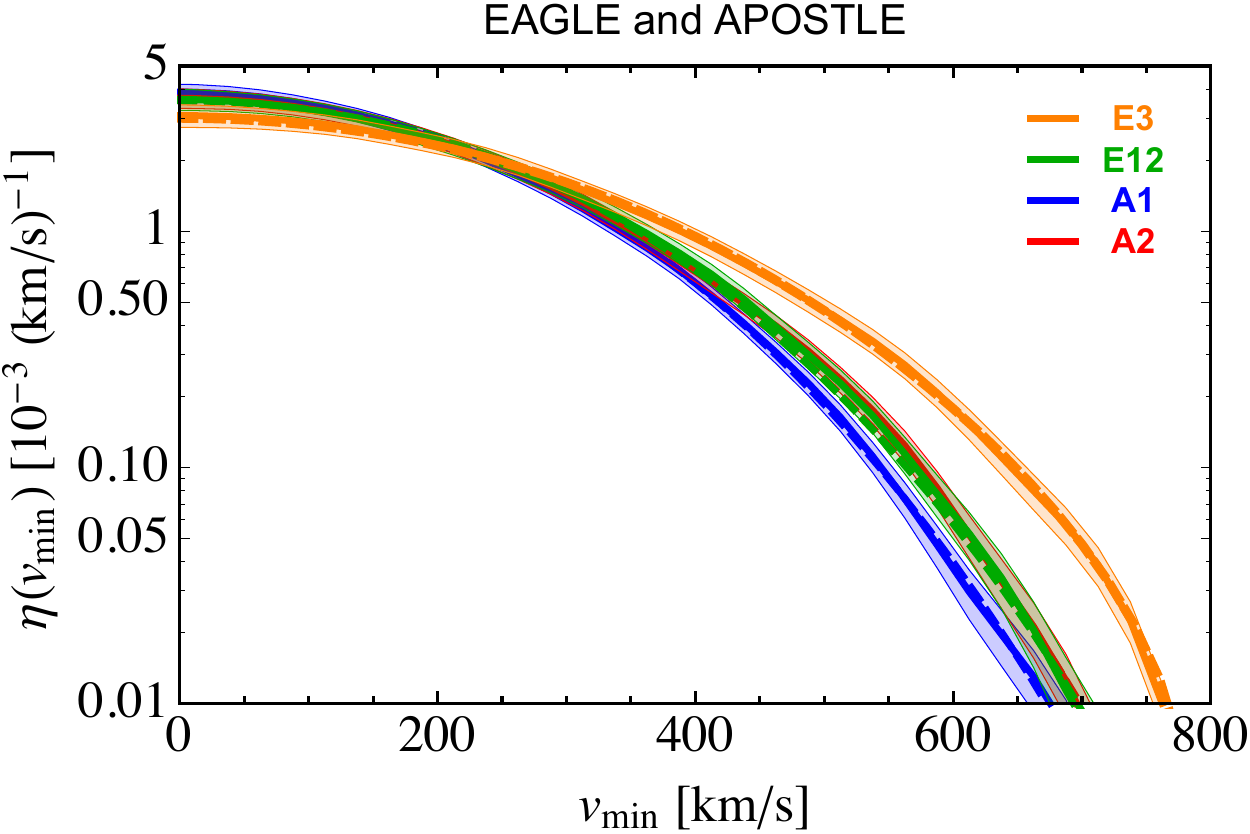}
\caption{Time averaged halo integrals for (a) haloes in different simulations (colored lines) compared to the SHM Maxwellian halo integral with peak speed of 220 km$/$s and escape speed of 544~km$/$s (dashed black line), and (b) the two haloes in the EAGLE HR simulation with DM speed distributions closest to (green) and farthest from (orange) the SHM Maxwellian and the two MW analogues in the APOSTLE IR simulation (blue and red). The solid lines and shaded regions show the mean halo integral and its 1$\sigma$ uncertainty, respectively. The dashed colored lines, which are very similar to the solid lines, show the halo integrals obtained from the best fit Maxwellian speed distribution for each halo (with matching colors). The curves in the left panel are obtained from Refs.~\cite{Kuhlen:2013tra, Bozorgnia:2016ogo, Kelso:2016qqj, Sloane:2016kyi}. The green curve in the left panel shows the average of the maximum and minimum halo integrals presented in Sloane {\it et al.}~\cite{Sloane:2016kyi}.
 \label{fig:eta}}
\end{figure}

The left panel of Fig.~\ref{fig:eta} shows the time averaged halo integrals (Eq.~\ref{eq:eta}) obtained from the DM velocity distributions of simulated haloes in different simulations as a function of the minimum WIMP speed, $v_{\rm min}$. To compare the results of different simulations with the SHM Maxwellian (with peak speed of 220~km$/$s and escape speed of 544~km$/$s), the halo integral for the halo which has the farthest DM speed distribution in the Galactic rest frame from the SHM (specified with a dashed black line) is shown only. The DM speed distributions for the same haloes are presented in the left panel of Fig.~\ref{fig:fv}. 

The right panel of Fig.~\ref{fig:eta} shows the halo integrals and their 1$\sigma$ uncertainties for the two MW analogues in the EAGLE HR simulation with DM speed distributions closest to (halo E12, shown in green) and farthest from (halo E3, shown in orange) the SHM Maxwellian and the two MW analogues in the APOSTLE IR simulation (haloes A1 and A2, shown in blue and red, respectively). The DM speed distributions for the same haloes  are shown in the right panel of Fig.~\ref{fig:fv}. The halo integrals obtained from the best fit Maxwellian speed distributions are shown as dashed curves with matching colors for each halo. To obtain the best fit Maxwellian halo integral for each halo, we boost the best fit Maxwellian speed distribution in the Galactic  rest frame by the local circular speed at 8 kpc for that halo\footnote{Notice that in Ref.~\cite{Bozorgnia:2016ogo} the best fit Maxwellian halo integrals were obtained by boosting the best fit Maxwellian speed distribution in the Galactic rest frame for each halo by its best fit Maxwellian peak speed, instead of boosting by its local circular speed. Since boosting by the local circular speed of each halo is more consistent with the simulation data, we obtain an even better fit to the halo integrals of the simulated haloes  (as shown in the right panel of Fig.~\ref{fig:eta}) compared to the results presented in Ref.~\cite{Bozorgnia:2016ogo}.}.  

Below we discuss how well the halo integrals of the simulated haloes fit the SHM or the best fit Maxwellian halo integral.

\subsubsection{Eris}

The time-averaged halo integral of the halo in Eris is compared to the SHM with peak speed of 220 km$/$s. At low and intermediate values of $v_{\rm min}$ there are only small differences between the Eris and SHM halo integrals. In particular, for $v_{\rm min}<180$~km$/$s, the halo integral in Eris is larger by up to 15\% compared to the SHM, but is suppressed at larger $v_{\rm min}$. The suppression of the halo integral at high $v_{\rm min}$ compared to the SHM may be due to the non-rotating DM density enhancement in the disk.

The tail of the halo integral in Eris is moved to higher DM speeds compared to ErisDark. This is due to the broadening of the DM speed distribution in Eris compared to ErisDark as a result of the halo contraction.

\subsubsection{EAGLE and APOSTLE}

As inferred from Fig.~\ref{fig:eta}, there is a significant halo-to-halo scatter in the halo integrals of the MW analogues in the EAGLE HR and APOSTLE IR simulations. The halo integrals obtained from the best fit Maxwellian speed distributions fall within the 1$\sigma$ uncertainty band of the halo integrals for all but one MW analogue, where there is only a very small deviation at large $v_{\rm min}$. On the other hand, the SHM Maxwellian with a peak speed of 220 km$/$s does not fit the halo integrals of all the MW analogues, especially at higher $v_{\rm min}$. This is due to the different peak speeds of the DM velocity distributions of the MW analogues compared to the SHM.

\begin{figure}[t]
\includegraphics[width=0.49\textwidth]{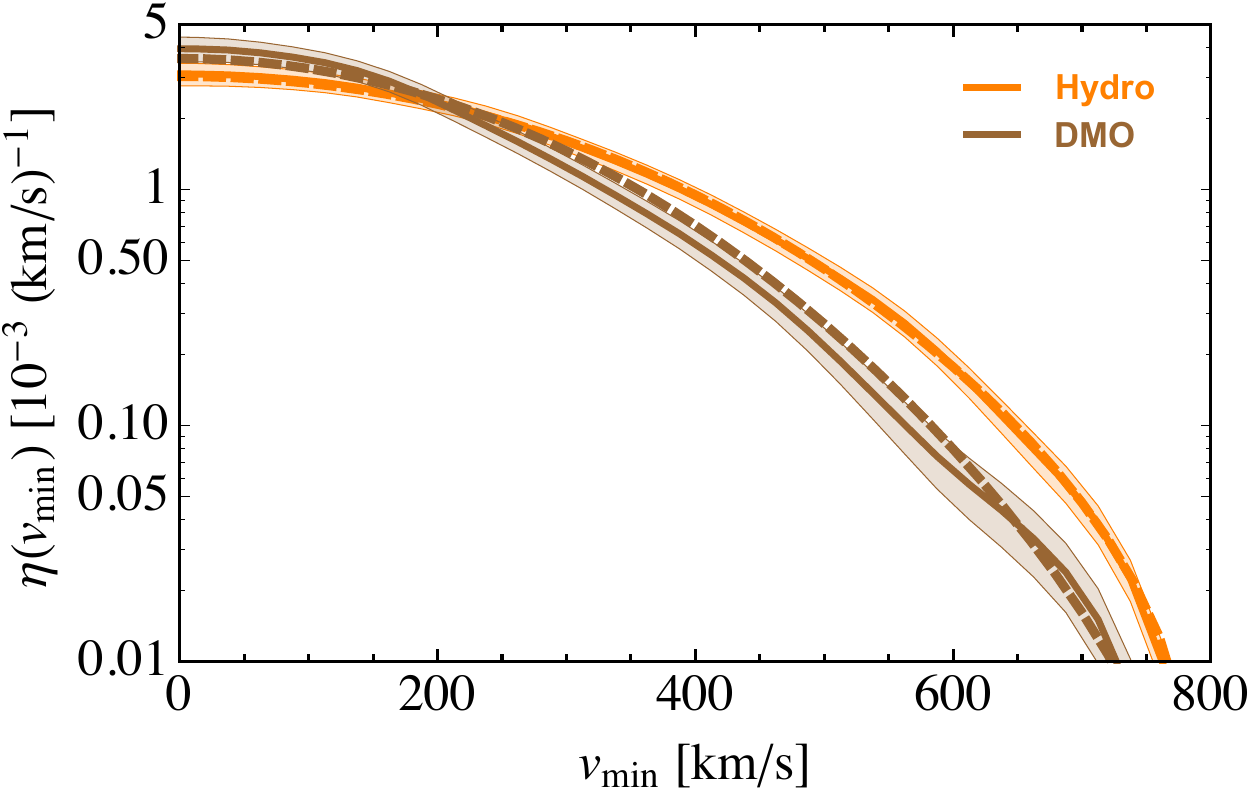}
\includegraphics[width=0.49\textwidth]{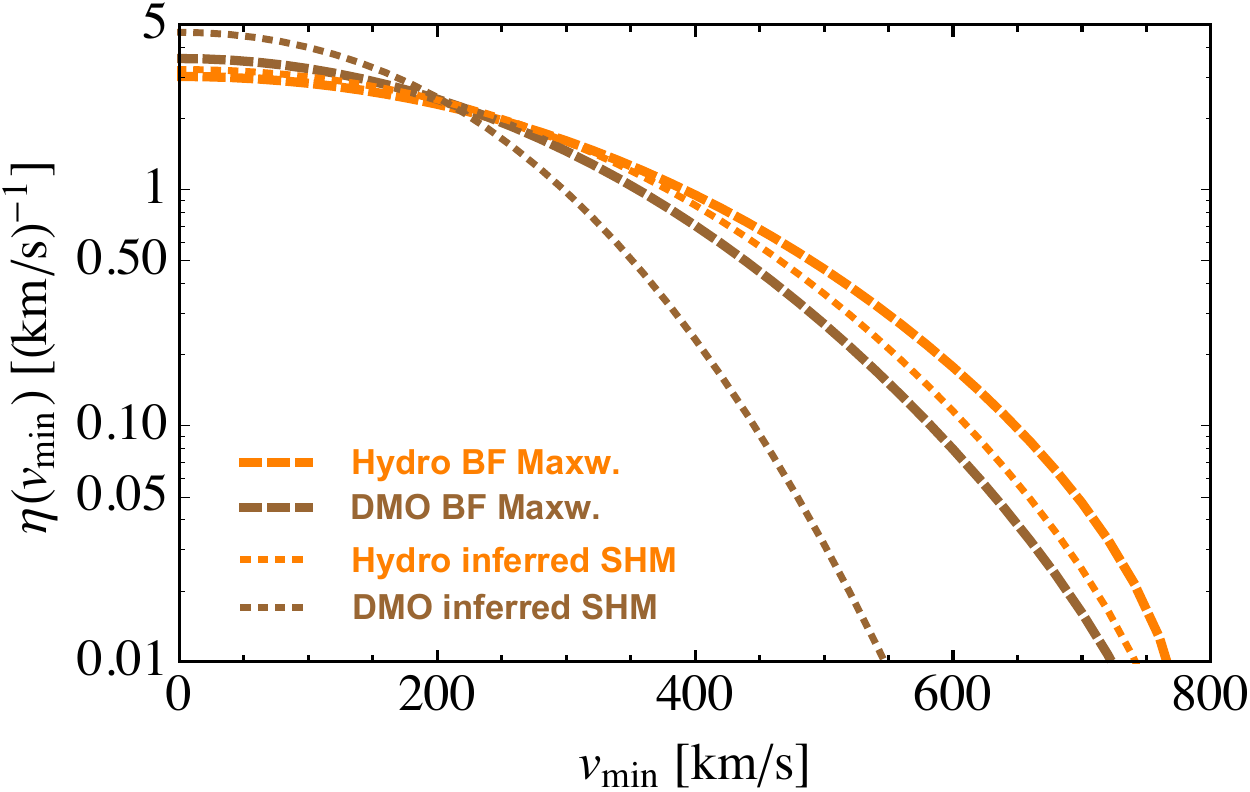}
\caption{(a) Comparison of the halo integrals for a MW analogue (halo E3) in the EAGLE HR hydrodynamical simulation (solid orange line and its 1$\sigma$ error band) and its DMO counterpart (solid brown line and its 1$\sigma$ error band) obtained from the DM speed distributions shown in Fig.~\ref{fig:HydroDMO}. Dashed lines specify the halo integrals obtained from the best fit Maxwellian distributions for the hydrodynamical (orange) and DMO (brown) haloes. (b) Comparison of the halo integrals obtained from the best fit Maxwellian (dashed lines, same as shown in panel (a)) and the inferred SHM  distributions (dotted lines) for the same MW analogue and its DMO counterpart shown in panel (a).
 \label{fig:etaDMOHydro}}
\end{figure}

In the left panel of Fig.~\ref{fig:etaDMOHydro} we show a comparison of the halo integrals of a MW analogue in the EAGLE HR hydrodynamical simulation and its DMO counterpart obtained from the DM speed distributions presented in Fig.~\ref{fig:HydroDMO}. Baryons significantly affect the velocity distribution and the halo integrals at the Solar radius, resulting in a shift of the tails of the halo integral towards higher minimum velocities compared to the DMO case. For both the hydrodynamical and DMO cases, the best fit Maxwellian halo integral is computed from the best fit Maxwellian DM speed distribution in the Galactic rest frame boosted by the local circular speed of the corresponding halo. The best fit Maxwellian halo integrals computed in this way fall within the 1$\sigma$ uncertainty band of the halo integrals of the simulated halo in both cases, although the fit is better in the hydrodynamical case.  

In the right panel of Fig.~\ref{fig:etaDMOHydro} we show a comparison of the best fit Maxwellian halo integrals (dashed lines, shown also in the left panel of the same figure) and the inferred SHM halo integrals (dotted lines), for each halo in the hydrodynamical and DMO cases.  The inferred SHM halo integrals are computed by assuming a Maxwellian DM speed distribution for each halo with a peak speed equal to the local circular speed of that halo. For the DMO case, the inferred SHM and the best fit Maxwellian halo integrals are significantly different. This is due to the large difference between the best fit Maxwellian peak speed of the DMO haloes and their local circular speed, as shown in Fig.~\ref{fig:vpeakvc} and discussed in Section~\ref{vpeakvc}. The inferred SHM, however, provides a better fit to the halo integral of the hydrodynamical halo compared to the DMO case. 

These conclusions remain the same for all the MW analogues in the EAGLE HR and APOSTLE IR simulations and their DMO counterparts. In general, halo integrals computed from the best fit Maxwellian speed distributions boosted by the local circular speed of each halo provide an excellent fit to the halo integrals of the MW analogues.

\subsubsection{MaGICC}

As seen in Fig.~6 of Ref.~\cite{Kelso:2016qqj}, the halo integrals of the two MW analogues in MaGICC fit well the halo integrals for the SHM inferred from the mass of each halo as well as those for the best fit Maxwellian with and without rotation. Notice that some deviations exist at high $v_{\rm min}$ compared to the inferred SHM, which may be due to the low particle counts in the simulations at high speeds~\cite{Kelso:2016qqj}. 

The inferred SHM provides a a very poor fit to the halo integral of the halo in the DMO simulation. However the best fit Maxwellian with and without rotation both provide good fits for the DMO case. Notice that these conclusions are the same as those reached for the EAGLE simulations.

\subsubsection{Sloane {\it et al.}}

Sloane {\it et al.} study the maximum and minimum halo integrals during a year. They find that the halo integrals of the simulated haloes are lower than the SHM (with peak speed of 220 km$/$s) at high velocities, and higher than the SHM at low velocities. However, the SHM halo integrals show an improved fit to the halo integrals of the simulated haloes in the hydrodynamic simulation compared to the DMO. From Fig.~3 of Ref.~\cite{Sloane:2016kyi} one can also notice that other than one of the selected haloes which has a prominent dark disk (h258), the best fit Maxwellian halo integrals show only small deviations at high $v_{\rm min}$ with respect to the halo integrals of the simulated haloes. 

In Fig.~\ref{fig:eta} we show the time-averaged halo integral for this same halo which has the farthest halo integral compared to the SHM, computed from the maximum and minimum halo integrals for h258 presented in Fig.~3 of Ref.~\cite{Sloane:2016kyi}.


\subsection{Dark matter parameters}

The DM distribution extracted from simulations could be used directly in the analysis of data from different direct detection experiments. 
We discuss how the allowed regions and exclusion limits set by direct detection experiments in the spin-independent cross section and DM mass change compared to the SHM for the MW-like haloes studied in Refs.~\cite{Bozorgnia:2016ogo}, \cite{Kelso:2016qqj} and \cite{Sloane:2016kyi} using the EAGLE HR, MaGICC, and Sloane {\it et al.} simulations. 
The other hydrodynamical simulations either do not perform an analysis of direct detection data, or the data used are outdated, and therefore we do not present them here.

For the purpose of comparing the predictions of different simulations for DM direct detection, we consider the positive signal from DAMA/LIBRA~\cite{Bernabei:2013xsa} (DAMA for short) and the null result from the LUX~\cite{Akerib:2013tjd} experiment. 
The DAMA experiment which measures the scintillation signal in their  NaI crystals has observed a 9.3$\sigma$ annual modulation signal over 14 annual cycles for the total exposure of 1.33 ton yr. 
The data on the DAMA annual modulation amplitude can be used to set the preferred DAMA regions in the $m_{\rm \chi}$ -- $\sigma_{\rm SI}$ plane. 
The LUX experiment is a dual phase (liquid $+$ gas) detector operating in the Sanford Underground Laboratory in South Dakota, and measures the ionization and scintillation signals. The latest results presented by the LUX experiment~\cite{Akerib:2016vxi} for an exposure of $3.35 \times 10^4$ kg day is consistent with background.

Notice that for the analysis of DAMA data, Refs.~\cite{Bozorgnia:2016ogo} and \cite{Kelso:2016qqj} use the latest DAMA results from Ref.~\cite{Bernabei:2013xsa}, while Ref.~\cite{Sloane:2016kyi} uses the slightly older DAMA data from Ref.~\cite{Bernabei:2010mq} with a total exposure of 1.17 ton yr. Notice also that Refs.~\cite{Bozorgnia:2016ogo}, \cite{Kelso:2016qqj} and \cite{Sloane:2016kyi} all use the 2014 LUX data~\cite{Akerib:2013tjd} which had an exposure of $\sim 10^4$ kg day. The LUX data has been updated since, but the conclusions discussed in \cite{Bozorgnia:2016ogo}, \cite{Kelso:2016qqj}, \cite{Sloane:2016kyi} and summarized here, remain the same. We refer the reader to Refs.~\cite{Bozorgnia:2016ogo}, \cite{Kelso:2016qqj} and \cite{Sloane:2016kyi} for the details of the analysis of direct detection data.

\begin{figure}[t]
\includegraphics[width=0.49\textwidth]{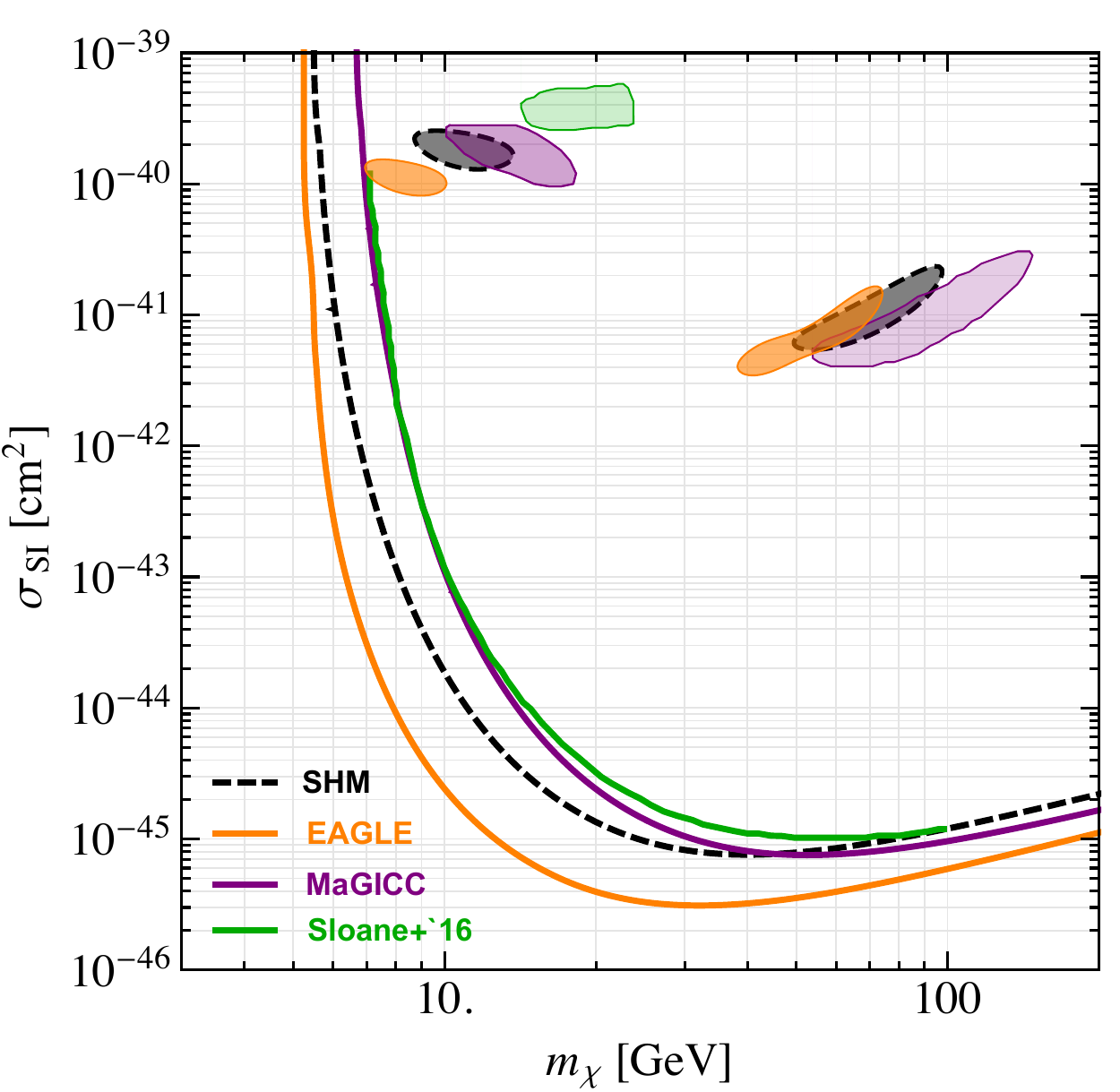}
\includegraphics[width=0.49\textwidth]{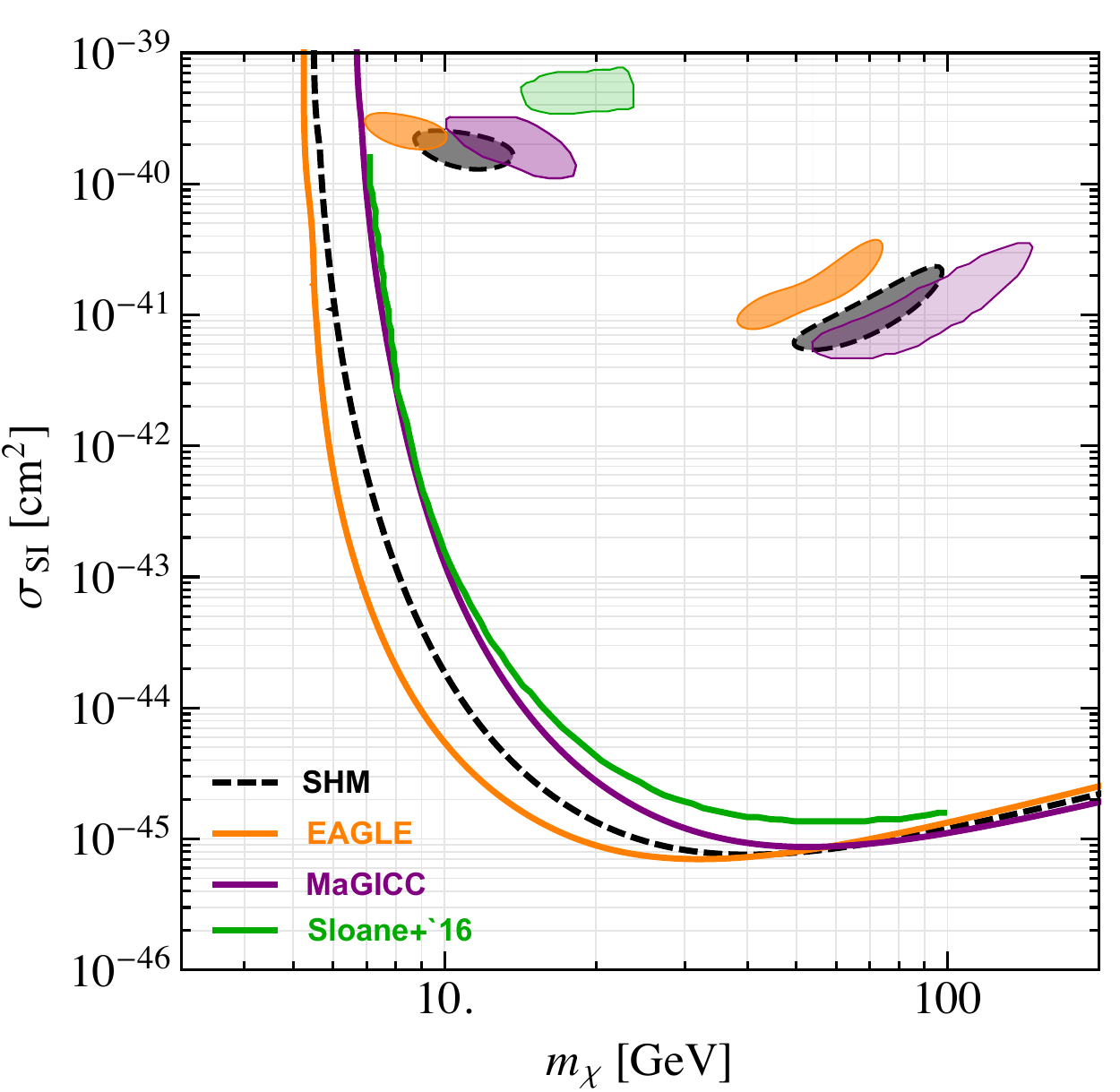}
\caption{LUX exclusion limit (at 90\% CL) and DAMA preferred regions (at 3$\sigma$) in the spin-independent DM-nucleon cross section and DM mass plane for a halo in the EAGLE HR (orange), MaGICC (purple), and Sloane {\it et al.} (green) simulations, as compared to the SHM Maxwellian (dashed black) with peak speed of 220 km$/$s. The local DM density is different for each halo in the left panel, whereas $\rho_\chi=0.3$~GeV$/$cm$^3$ for all haloes in the right panel. The curves in the left panel are obtained from Refs.~\cite{Bozorgnia:2016ogo, Kelso:2016qqj, Sloane:2016kyi}, and rescaled for the local DM density of 0.3~GeV$/$cm$^3$ to obtain the curves in the right panel.
\label{Fig:ExcLim}}
\end{figure}

The left panel of Fig.~\ref{Fig:ExcLim}  shows the LUX exclusion limit (at 90\% CL) and the DAMA allowed region (at 3$\sigma$) for a MW-like halo from the EAGLE HR, MaGICC\footnote{Notice that the LUX exclusion limit for the MW analogue in MaGICC presented here is weaker at low DM masses than those presented in Fig.~9 of Ref.~\cite{Kelso:2016qqj}. The reason is that Ref.~\cite{Kelso:2016qqj} does not use the same lower energy bound of 3 keV that LUX uses when computing the exclusion limits~\cite{KelsoPrivCom}. Therefore, instead of reading off the LUX exclusion limit for the MaGICC MW-like halo from Ref.~\cite{Kelso:2016qqj}, we compute it from the halo integral presented in Fig.~6 of Ref.~\cite{Kelso:2016qqj} using the 3 keV LUX energy threshold.} and Sloane {\it et al.} simulations, which has a local DM speed distribution farthest from the SHM Maxwellian with a peak speed of 220 km$/$s. The local DM speed distributions and halo integrals for the same haloes are shown in the left panels of Figs.~\ref{fig:fv} and \ref{fig:eta}, respectively. The local DM density of the haloes in the EAGLE HR (halo E3) and MaGICC (halo g1536) simulations is 0.68 and 0.346 GeV$/$cm$^3$, respectively, which is used to obtain the exclusion limits and preferred regions in the left panel of Fig.~\ref{Fig:ExcLim}. For the Sloane {\it et al.} halo, the local DM density is set to 0.4 GeV$/$cm$^3$~\cite{Sloane:2016kyi}. 

In order to solely show the effect of the DM velocity distribution on direct detection results, in the right panel of Fig.~\ref{Fig:ExcLim} we show the results for the same haloes in the left panel, but setting the local DM density to 0.3 GeV$/$cm$^3$ for all haloes.

One can see from Fig.~~\ref{Fig:ExcLim}  that the largest difference between the exclusion limits and allowed regions with respect to the SHM at all DM masses is due to variation of the local DM density of the simulated haloes with respect to the SHM. The shift at low DM masses is the result of the different high velocity tail of the halo integrals of the simulated haloes with respect to the SHM. 

\section{Comments on non-standard interactions}
\label{Nonstandard}

The direct detection implications discussed so far in this review hold for the case of standard spin-independent and spin-dependent DM-nucleus interactions, where the energy differential cross section which enters in the direct detection event rate is proportional to $v^{-2}$ (Eq.~\ref{eq:dsigmadE} for the spin-independent case). This leads to event rates proportional to the halo integral, $\eta(v_{\rm min}, t)$, as given in Eqs.~\ref{eq:Reta} and \ref{eq:eta}. The DM-nucleus interaction can, however, be more complicated. For non-standard interactions, the differential  cross section can have a different dependence on the relative velocity between the DM and the nucleus, and also depend on the exchanged momentum. The classification of all possible DM-nucleus interactions has been performed in the non-relativistic limit~\cite{Fan:2010gt, Fitzpatrick:2012ix, Fitzpatrick:2012ib}. For a very general set of non-relativistic effective operators,  the energy differential cross section  can be written as a linear combination of a velocity-dependent term proportional to $v^{-2}$ and a velocity-independent term~\cite{DelNobile:2013sia, Kahlhoefer:2016eds}.  The former gives rise to the usual halo integral, $\eta(v_{\rm min}, t)$, while the velocity-independent term gives rise to a new velocity integral in the form of 
\begin{equation}\label{eq:h}
h(v_{\rm min}, t)=\int_{v>v_{\rm min}} d^3 v~v~ f_{\rm det}({\bf v},t).
\end{equation} 
As a result, direct detection event rates will be a sum of two terms, one proportional to the usual halo integral, $\eta(v_{\rm min}, t)$, and the other proportional to $h(v_{\rm min}, t)$.  An example  of  this linear combination of the two different velocity dependences is the  case of the magnetic dipole DM, which the DM is assumed to be a fermion interacting via a magnetic dipole moment with the nucleus.

\begin{figure}[t]
\centerline{\includegraphics[width=0.6\textwidth]{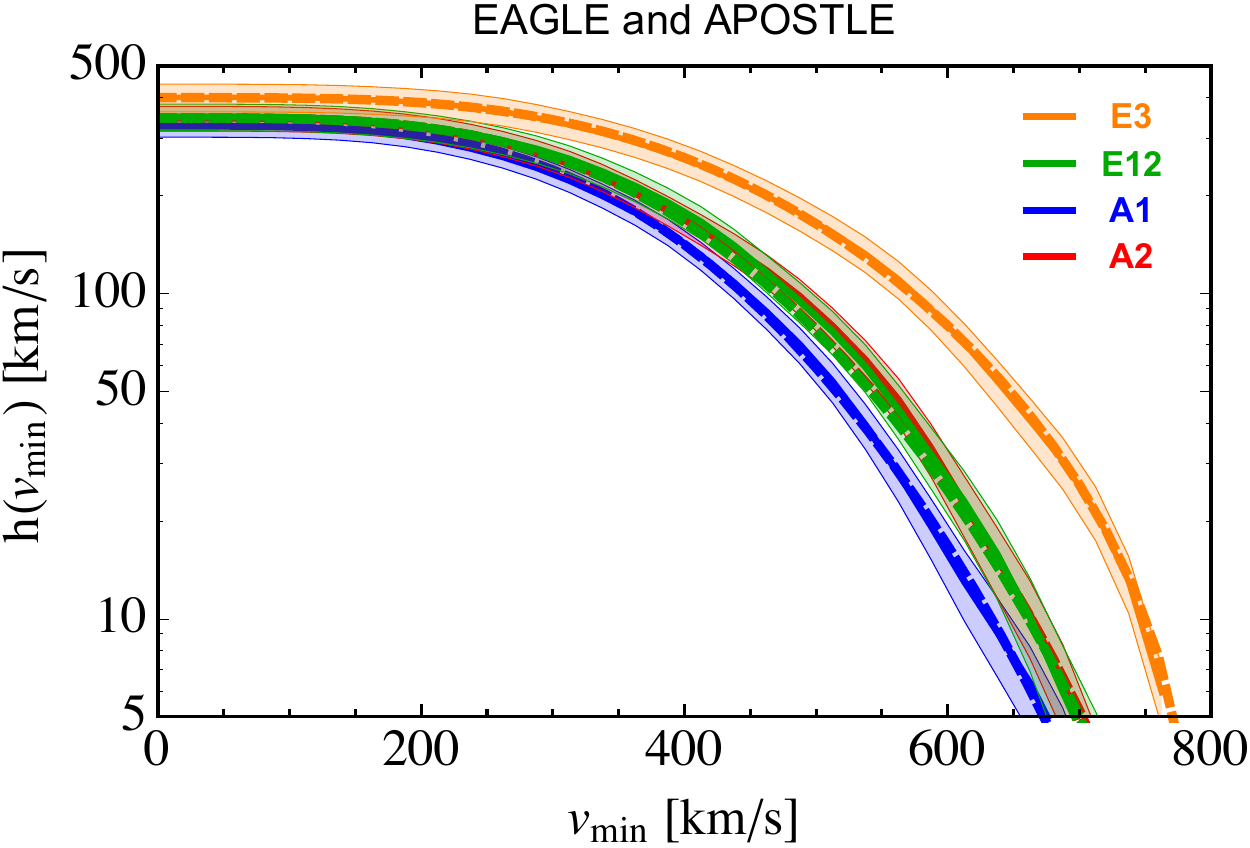}}
\caption{Same as the right panel of Fig.~\ref{fig:eta}, but showing the time-averaged velocity integral, $h(v_{\rm min})$ (given in Eq.~\ref{eq:h}).
 \label{fig:h}}
\end{figure}

We extract the velocity integral, $h(v_{\rm min}, t)$ from the local DM distributions in the EAGLE HR and APOSTLE IR simulations and compare them with $h(v_{\rm min}, t)$ obtained from the best fit Maxwellian speed distributions in the Galactic rest frame and boosted by the local circular speed of each halo. Figure~\ref{fig:h} shows the velocity integrals for the same four haloes shown in the right panel of Fig.~\ref{fig:eta}. The conclusions are similar to those obtained for the usual halo integral, $\eta(v_{\rm min})$. Namely, for all but one MW analogue, the best fit Maxwellian $h(v_{\rm min})$ fall within the 1$\sigma$ uncertainty band of the time-averaged $h(v_{\rm min})$ of the simulated haloes. For one halo, only a very small deviation at large $v_{\rm min}$ exists between the function $h(v_{\rm min})$ extracted from the simulated halo and the best fit Maxwellian distribution.

\section{Discussion and Conclusions}
\label{conclusions}

We have reviewed the status of hydrodynamical simulations and their implications for DM direct detection. Despite the diversity of numerical approaches, and of the criteria adopted to identify MW-like galaxies, a number of common trends can be identified, and interesting conclusions can be drawn in particular on the impact of baryonic physics on the local dark matter velocity distribution; the presence of dark disks; and on the implications for the direct detection of dark matter with standard and non-standard interactions.
We summarize here the main conclusions:

\begin{itemize}

\item {\bf Identification of Milky Way-like galaxies:} The criteria used to select simulated galaxies which resemble the MW are widely different among different simulation groups (see Table~\ref{tab:MW-like}). We recommend the prescription introduced in Ref. ~\cite{Bozorgnia:2016ogo}, based on a selection of haloes in the mass range $5\times10^{11}<M_{200}/\Msun<1\times10^{14}$; with a distribution of dark matter and baryons fitting the observed MW rotation curve~\cite{Iocco:2015xga,Pato:2017yai}; and with stellar mass in the range inferred from observations $4.5 \times10^{10}<M_{*}/\Msun<8.3 \times10^{10}$~\cite{McMillan:2011wd}.

\item {\bf Local DM density and halo shape:} The definition of the local DM density varies among simulations, but it was always found consistent with the range inferred from local and global observations $\rho_\chi = (0.2 - 0.8)$ GeV$/$cm$^3$~\cite{Read:2014qva, Catena:2009mf, Weber:2009pt, Salucci:2010qr, McMillan:2011wd, Garbari:2011dh, Iocco:2011jz, Garbari:2011tv, Bovy:2012tw, Zhang:2012rsb, Bovy:2013raa,Pato:2015dua, Silverwood:2015hxa, Huang:2016, McMillan:2016}. Although the shape of the inner DM halo varies in different simulations, baryons always tend to make the haloes more spherical\footnote{ This conclusion could not be verified in the case of Ling {\it et al.}, who do not have a DMO case, and Sloane {\it et al.}, who do not study the shape of DM haloes.}. The sphericity (minor to major axis ratio) of the inner haloes in the selected MW-like galaxies in different simulations studied here are in the range of $s=[0.69, 0.95]$.

\item {\bf Local DM speed distribution:} In all the simulations analyzed above, including baryons results, as expected, in shifting the peak of the local DM speed distributions to higher speeds, with respect to the DMO case (except in Ling {\it et al.}, which do not have a DMO case). This is due to the dissipative nature of baryons which fall into the center of the MW halo and make the gravitational potential deeper in the inner halo. Including baryons also appear to make the local DM speed distributions more Maxwellian in most cases. One notable exception is however the NIHAO simulation, where the DM speed distributions are less Maxwellian when baryons are included, possibly due to strong baryonic feedback.

\item {\bf Dark disks:} None of the simulations we analyzed point towards the existence of a sizable dark disk that can significantly impact direct detection experiments. Ling {\it et al.} find a 25\% fraction for the dark disk, whereas the Eris simulation finds a 3.2--9.1\% dark disk fraction depending on how this fraction is computed. No dark disks are found in the NIHAO and MaGICC simulations, and dark disks are rare in the EAGLE and APOSTLE simulations. Some of the galaxies studied in Sloane {\it et al.} have a dark disk. The cases where dark disks appear more prominent arise from simulations in which a large satellite merged with the MW in the recent past, a circumstance robustly excluded from MW kinematical data.

\item {\bf Halo integrals:} The halo integrals of simulated MW-like galaxies in the EAGLE and APOSTLE, MaGICC, and Sloane {\it et al.} simulations are similar to those obtained from the best fit Maxwellian velocity distributions. 
The only exception is the halo integral of one simulated galaxy in Sloane {\it et al.} which shows some discrepancy with respect to the best fit Maxwellian halo integral, due to the presence of a prominent dark disk (see discussion of dark disks above).
We have also checked the velocity integral (Eq.~\ref{eq:h}), which arises from the velocity-independent term in the energy differential cross section for non-standard interactions, and found again that for the MW analogues in EAGLE and APOSTLE, the best fit Maxwellian velocity integral provides an excellent fit to the velocity integral obtained from the simulations.

\item {\bf Implications for direct detection:} For the analysis of direct detection results we recommend the adoption of a Maxwell-Boltzmann velocity distribution, with a peak speed, $v_{\rm peak}$, constrained by hydrodynamical simulations of MW-like galaxies, and an independent local circular speed, $v_c$, constrained by observations or simulations. 
A better selection of  MW-like galaxies in numerical simulations, along the lines discussed above, will substantially reduce the astrophysical uncertainties in the exclusion plots and allowed regions set by direct DM experiments.
\end{itemize}

\section*{Acknowledgments}

We thank Mark Lovell and Matthieu Schaller for useful discussions, and providing valuable feedback on this review. We especially thank Chris Kelso for detailed discussions on the MaGICC simulation results. N.B. would like to express a special thanks to the Mainz Institute for Theoretical Physics (MITP) for its hospitality and support during the 2016 workshop ``Dark Matter in the Milky Way". G.B. (P.I.) and N.B.~acknowledge support from the European Research Council through the ERC starting grant WIMPs Kairos.

\bibliographystyle{JHEP.bst}
\bibliography{refs}

\end{document}